\documentclass{aastex}

\usepackage{txfonts}
\usepackage{natbib}
\usepackage{graphicx}
\usepackage{multirow}

\newcommand \x {X-ray}

\newcommand \pds {PDS\,456}
\newcommand{\uv}{{\small UV}}
\newcommand{\chandra}{{\it Chandra}}
\newcommand{\xmm}{{\small \it XMM-Newton}}

\newcommand{\asca}{{\small \it ASCA}}
\newcommand{\xte}{{\small \it RXTE}}
\newcommand{\epic}{{\small EPIC}}

\newcommand{\pn}{{\small EPIC~PN}}
\newcommand{\mos}{{\small MOS}}
\newcommand{\rgs}{{\small RGS}}
\newcommand{\hetg}{{\small HETGS}}

\newcommand{\suz}{{\small \it Suzaku}}
\newcommand{\xstar}{{\small \it XSTAR}}
\newcommand{\xspec}{{\small \it Xspec}}
\newcommand{\mcg}{MCG\,-6-30-15}

\newcommand{\cgsflux}{erg\,s$^{-1}$\,cm$^{-2}$}
\newcommand  \kms      {\ifmmode \mathrm{km\,s}^{-1} \else km\,s$^{-1}$\fi}
\newcommand  \cmii     {\hbox{cm$^{-2}$}}
\def\lsim{\mathrel{\rlap{\lower4pt\hbox{$\sim$}}
    \raise1pt\hbox{$<$}}}                
\def\gsim{\mathrel{\rlap{\lower4pt\hbox{$\sim$}}
    \raise1pt\hbox{$>$}}}                

\shorttitle{Variability of PDS 456}
\shortauthors{Behar et al.}

\begin{document}

\title{Remarkable Spectral Variability of PDS 456}

\author{Ehud Behar \altaffilmark{1,2}, 
Shai Kaspi \altaffilmark{3},
James Reeves \altaffilmark{4}, 
T.~J. Turner  \altaffilmark{5},
Richard Mushotzky \altaffilmark{1} 
\& Paul T. O'Brien \altaffilmark{6}}

\altaffiltext{1}{ Code 662, NASA/Goddard Space Flight Center, Greenbelt, MD 20771, USA}
\altaffiltext{2}{ Senior NPP Fellow on leave from the Technion, Israel}
\altaffiltext{3}{ Department of Physics, Technion, Haifa 32000, Israel}
\altaffiltext{4}{ School of Physics and Geographical Sciences, Keele University, Keele, Staffordshire ST5 5BG, UK}
\altaffiltext{5}{ Department of Physics, University of Maryland Baltimore County, 1000 Hilltop Circle, Baltimore, MD 21250, USA}
\altaffiltext{6}{ Department of Physics and Astronomy, University of Leicester, Leicester LE1 7RH, UK}


\begin{abstract} 
We report on the highest to date signal-to-noise-ratio X-ray spectrum of the luminous quasar \pds, 
as obtained during two \xmm\ orbits in September 2007. 
The present spectrum is considerably different
from several previous \x\ spectra recorded for \pds\ since 1998.
The ultra-high-velocity outflow seen as recently as February 2007 is not detected in absorption.
Conversely, a significant reflection component is detected ($\Delta \chi ^2 = 313$ compared to a simple absorbed power law).
The reflection model suggests the reflecting medium may be outflowing at a velocity $v/c = -0.06 \pm 0.02$ ($\Delta \chi ^2 = 28$ compared to $v/c = 0$).
The present spectrum is analyzed in the context of the previous ones
in an attempt to understand all spectra within the framework of a single model.
We examine whether an outflow with variable partial covering of the \x\ source along the line of sight that also reflects the source from other lines of sight can explain the dramatic variations in the broad-band spectral curvature of \pds.
It is established that absorption plays a major role in shaping the spectrum of other epochs, 
while the 2007 \xmm\ spectrum is dominated by reflection, and the coverage of the source by the putative outflow is small ($< 20\%$).


\end{abstract}


\section{Introduction}
\pds\ was discovered as a radio-quiet luminous quasar in the course of the Pico dos Dias survey (PDS) 
of bright IRAS sources \citep{torres97}.
The optical spectrum of \pds\ features broad emission lines and a redenning-corrected luminosity 30\% higher than that of 3C\,273 \citep{simpson99}.
The high luminosity of \pds\ provides a low limit of $10^9$~M$_\odot$ on the black hole mass, just from the Eddington limit. 
Rapid \x\ variability of \pds\ was first noticed with the \xte\ satellite in 1998 by \citet{reeves00}, who observed the 2 -- 10 keV flux double in 17~ks, which is rather uncommon for such a luminous quasar and indicative of an unusually compact source of a few gravitational radii. 

A few years later, \citet{reeves03} reported an ultra high velocity outflow from \pds\ based on blue-shifted Fe K-shell absorption features detected in a 40~ks \xmm\ observation from 2001.
Extremely blue-shifted Fe-K absorption has been reported for other quasars as well 
\citep[e.g.,][]{chartas02, chartas03, poundsPG12}, 
Obviously, for a confirmed detection of an outflow and for conclusive velocity measurements,
it is most desirable to observe as many resolved lines as possible, all shifted by the same velocity.
This is most difficult, since in many objects the absorption is detected primarily from the most highly ionized species of Fe,
whose lines occur at energies where the sensitivity of current detectors drops sharply, and in ionization regimes that preclude absorption by all other species that are too (fully) ionized to produce lines.
The difficulty is then augmented by the rapid spectral variability of these outflows \citep{kaspi05, chartas07, reeves08}, which imposes fundamental limits on the signal-to-noise-ratio (S/N) of spectra and deems data stacking from separate observations almost useless.
All these have raised questions regarding the statistical robustness of some, but probably not all, of the blue-shifted absorption detections \citep{vaughan08}.
The only way to substantiate and further study these high velocity outflows is with long exposures and  high S/N spectra, such as the present one.

\pds\ stands out as one of the least controversial ultra high velocity outflows.
Not only does it have multiple Fe-K absorption detections over the years \citep{reeves00, reeves03, reeves09}, but a 2001 \xmm\ observation revealed an absorption feature in the soft \x\ band that was measured at high-resolution with the Reflection Grating Spectrometers (\rgs).
Indeed, the \rgs\ spectrum shows a deep and broad absorption trough between 13--14~\AA\ (in the observed frame), interpreted by \citet{reeves03} as due to a $\sim$50,000~\kms\ outflow that also produces the Fe-K absorption edge.
Blue-shifted Ly\,$\alpha$ absorption indicates an appreciably slower outflow moving at 14,000 -- 24,000 \kms\ \citep[][]{obrien05}.
Despite its moderate redshift of $z$ = 0.184, \pds\ is also the brightest among the targets suspected of harboring an ultra high velocity outflow.
With all this in mind, we obtained a deep \xmm\ exposure of \pds, which is reported in this paper.
\xmm\ has the largest effective area of all current \x\ telescopes and thus provides the best chance to observe several absorption lines from the same outflow component with the gratings.

In the meantime, a \suz\ observation of \pds\ provided another high-quality (though low-resolution) spectrum of \pds\ in which Fe-K line shifts of 100,000 \kms\ were detected at high S/N, but with no obvious soft \x\ line absorption \citep{reeves09}. This adds to the mounting evidence of blue-shifted absorption detected in \pds\ above 8~keV.  An exciting discovery in the \suz\ observation is the significant emission at high energies above 15~keV, much in excess of the extrapolated soft \x\ spectrum.
This is yet another manifestation of the unique spectral behavior exhibited by this intriguing quasar.
In the present paper, after analyzing the 2007 observations and studying the short term \x\ variability within those two \xmm\ orbits  in Section~\ref{sec:xmm07}, in Section~\ref{sec:partabs} we also study the spectral variability of \pds\ between different epochs, and on time scales of months and years, over the past decade. 

\section{Observations}
\label{datareduction}

\subsection{Data Reduction and Light Curves}

In Table~\ref{obs_log} we list the log of the 2007 \xmm\ observations of \pds\ as well as the earlier \xmm\ observation from 2001, and then the other observations of \pds\ that are used later in this paper.
\xmm\ data were reduced using the Science Analysis Software (SAS - ver. 7.1.0) in the standard processing chains as described in the ABC Guide to \xmm\ Data Analysis\footnote{See http://heasarc.gsfc.nasa.gov/docs/xmm/abc},
and using the most updated calibration files for 2009, June 15.
Both single and double events were retained and no pile up was detected.
1.5~ks were excluded from the orbit-1421 data set due to high background. 

\begin{figure}
\begin{center}
\vglue0.0mm
\includegraphics[width=0.65\columnwidth]{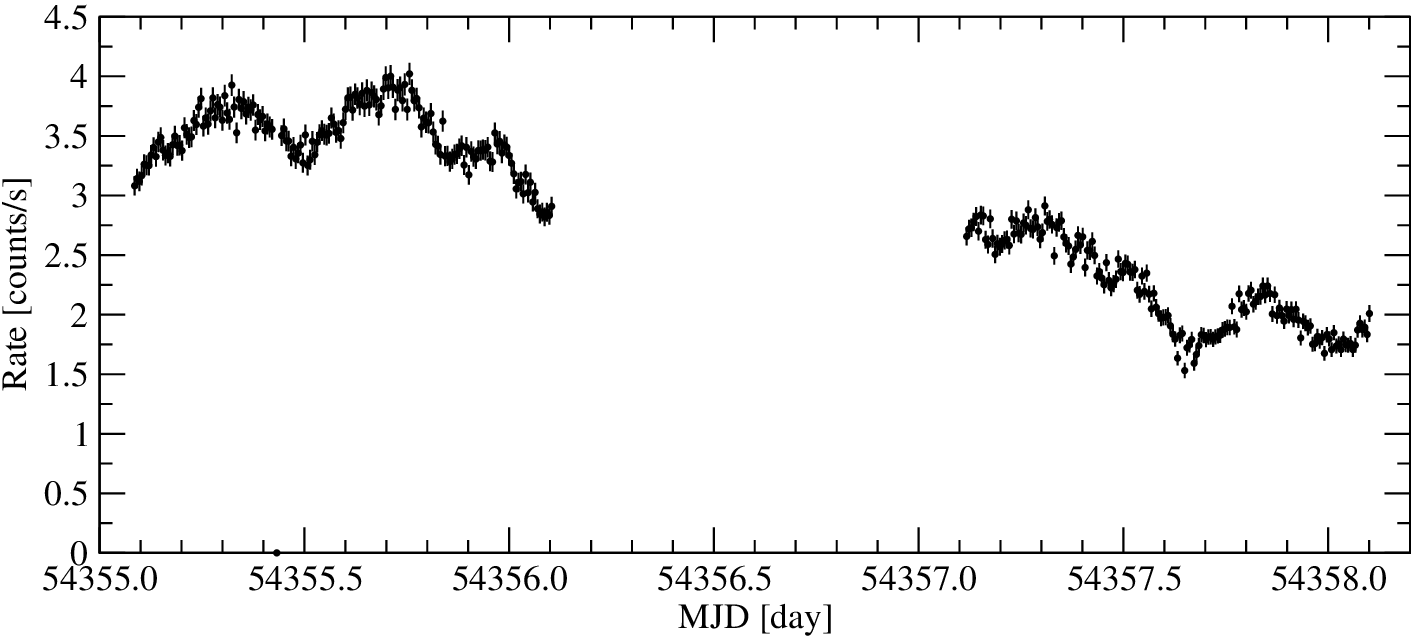} 
\caption{ \pn\ light curve of PDS 456 from two consecutive \xmm\ orbits during 2007, Sep. 12 -- 15.
Data are binned to 500~s and include the 0.2 -- 15 keV photon energy range.}
\label{fig:lc}
\end{center}
\end{figure}

The light curve of the source during orbits 1421 and 1422 is shown in Fig.~\ref{fig:lc}.
The source flux diminishes gradually during the second orbit to approximately half its initial flux level.
The typical time for variability in Fig.~\ref{fig:lc} is $\sim$30\,ks, 
similar to that reported by \citet{reeves02}, and which
corresponds to the light crossing time of only a few gravitational radii.
It should be noted that despite the significant \x\ variability of \pds, it is rather steady in longer wavebands. 
The optical monitor (OM) on board \xmm\ recorded a flux range of (2.4 -- 2.5) $\times 10^{-15}$ \cgsflux\ (1.10 -- 1.14 cts\,s$^{-1}$ during orbit 1421) with the UVM2 filter around 2300~\AA\  and a range of only (5.75 -- 5.85) $\times 10^{-15}$ \cgsflux\ (12.1 -- 12.3 cts\,s$^{-1}$  during orbit 1422) with the UVW1 filter around 2900~\AA. 
These can be further compared with the archival HST/STIS measured flux of  3 $\times 10^{-15}$ \cgsflux\  at 2300~\AA\ and 9.5 $\times 10^{-15}$ \cgsflux\  at 2900~\AA\ on 2000, May 14. 
All fluxes are quoted before correcting for reddening.
Optical monitoring in the V band carried out at the Wise Observatory every two
weeks over half a year reveals an increase in flux from 2007 April to
July of only 0.05 mag, and then a decrease in flux from 2007 July to October
of 0.07 mag.
Thus, while the X-ray flux varies by a few tens of percent
over a period of a day, the \uv\ flux is constant to within a few percent.
At much longer time scales, perhaps even years but not well monitored,
the \uv\ flux can also vary by a factor of two.
The optical flux varies by only a few percent over several months. 
This is not unusual, as quasar monitoring often shows an increase of variability with frequency 
from optical, through \uv, and up to the \x\ band \citep{edelson96}.

\subsection{Background}
\label{bg}

In order to obtain the \x\ spectra, we extracted the source photons from the \pn\ and \mos\ cameras in a circular region of radius 40\arcsec\ and grouped the data to bins with minimum 25 counts per bin.
We use the standard SAS background files for each observation, where the background spectrum is extracted from an annulus region centered around the source of 1\arcmin ~to 3\arcmin, from which 1\arcmin\ circular regions around field point sources are excluded. 
The background for \pn\ requires special caution as it can have strong instrumental, narrow emission lines at energies around 1.5 and 8 keV, which could introduce by subtraction artificial absorption lines.
The background during the 2007 observations was relatively high, but we verified that the background spectra we use have no such lines. 

\begin{figure}
\begin{center}
\includegraphics[width=0.75\columnwidth, angle=0]{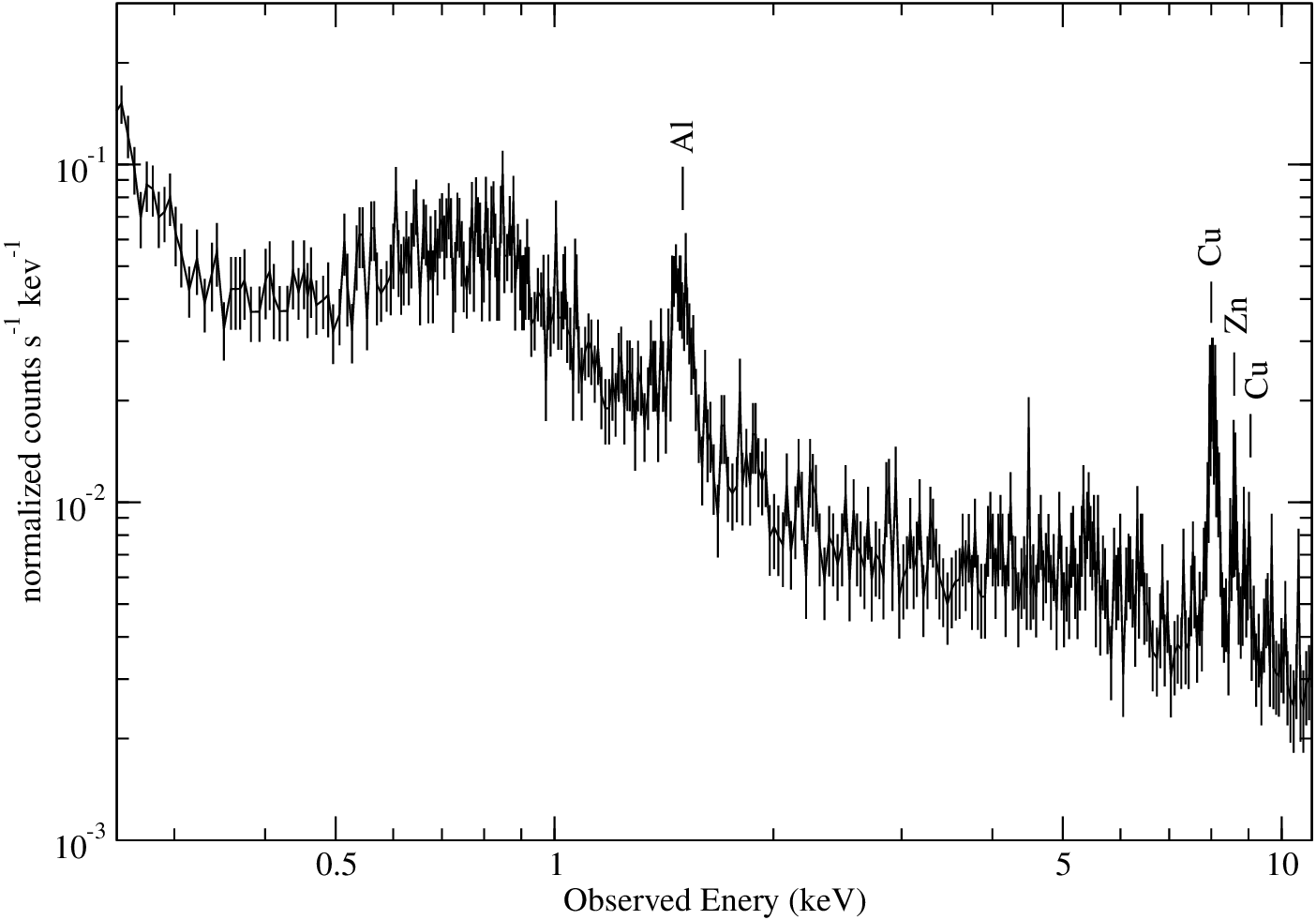} 
\caption{Typical off-source \pn\ background spectrum from the \xmm\ observation of \pds\ during orbit 1421.
}
\label{fig:bg} 
\end{center}
\end{figure}

We measured the background in various off-source locations on the \pn\ and \mos\ CCDs.
A sample background spectrum is shown in Fig.~\ref{fig:bg}.
Indeed, instrumental emission lines appear on the detector with varying strengths
as has been documented in the \xmm\ Users'
Handbook\footnote{http://heasarc.gsfc.nasa.gov/docs/xmm/uhb/node35.html}.
\citet{freyberg04} investigated the intrinsic instrument background in the \epic\
detectors and found that it is composed of a continuum and X-ray fluorescence lines,
most prominent are those at 1.5, 7.7, 8.0, 8.6, and 17.4 keV, which are due to K$\alpha$
line emission of Al, Ni, Cu, Zn, and Mo, respectively. 
The lines show strong spatial inhomogeneities 
correlated with the structures of the electronics board mounted below the \pn\ CCDs.
\citet{freyberg04} also found that the very central region of the camera where typically the primary source is located is free of this instrumental background.
Therefore, our use of line free background spectra is justified.

We also examined the background in the archival \xmm\ 2001 observation.
For this observation as well, we use the standard 1\arcmin ~to 3\arcmin\ annulus.
The background in the 2001 observation is much lower and its fluorescence lines 
much less prominent.
More precisely, at the edge of the CCD the fluorescence lines are strong,
but they get much weaker close to the center of the CCD, as expected \citep{freyberg04}. 
Thus, the spectral turnover detected around 8 keV in the 2001 spectrum of \pds\ \citep{reeves03} is
secure and not significantly influenced by the background.

\section{2007 \xmm\ Campaign}
\label{sec:xmm07}

\begin{figure}[pt!]
\begin{center}
{\includegraphics[angle=-90,width=9cm]{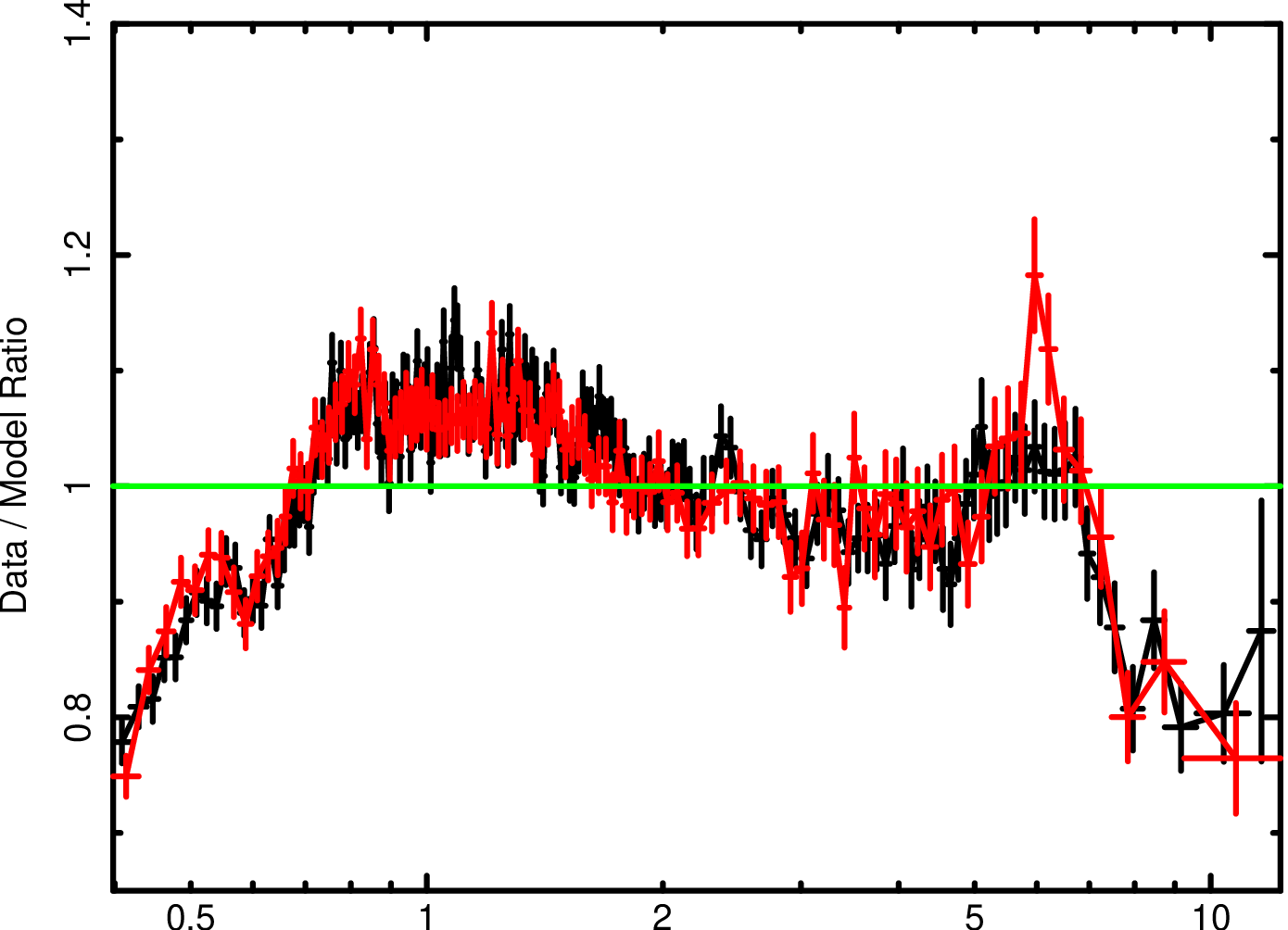}}
{\includegraphics[angle=-90,width=9cm]{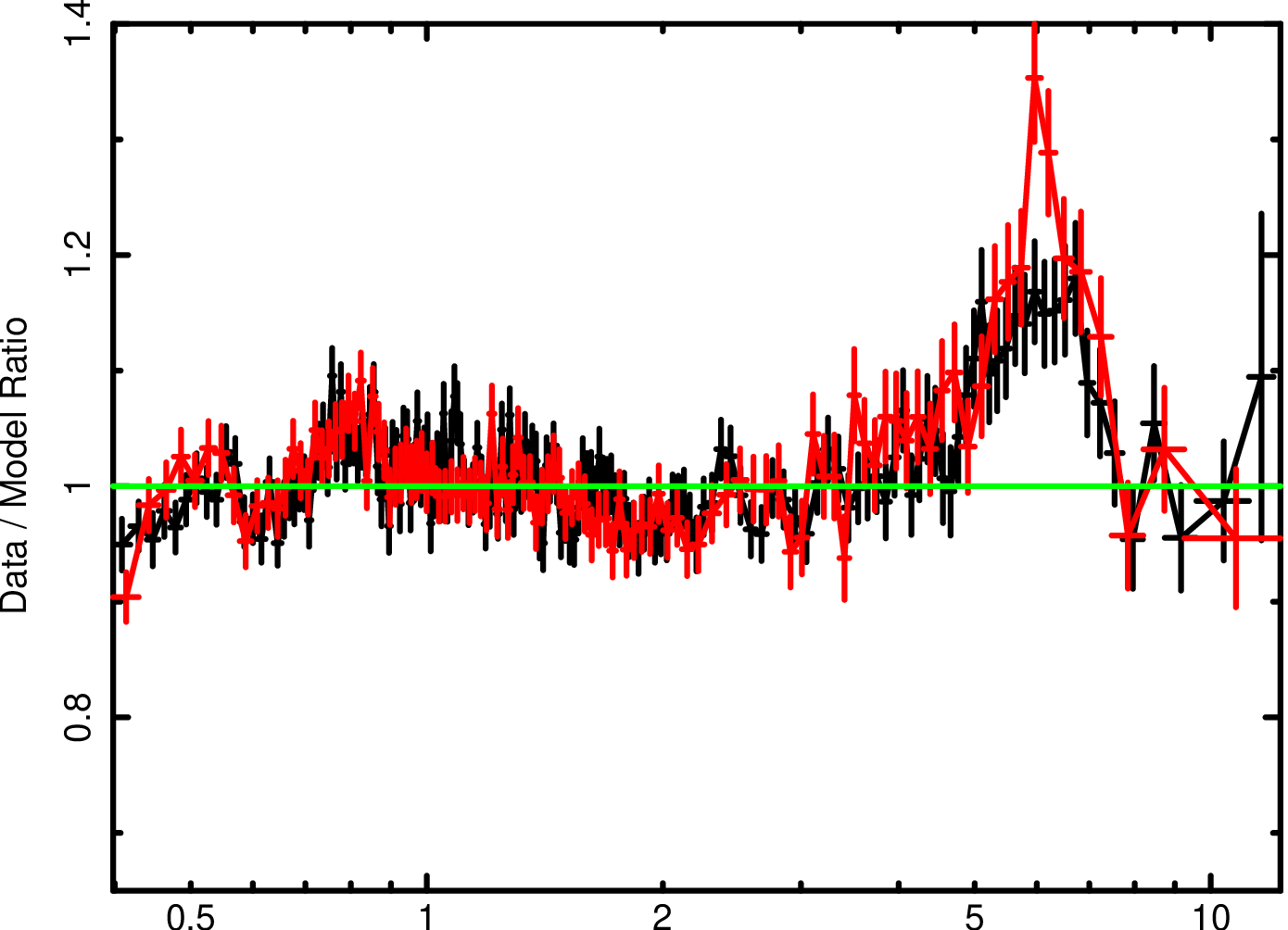}}
{\includegraphics[angle=-90,width=9cm]{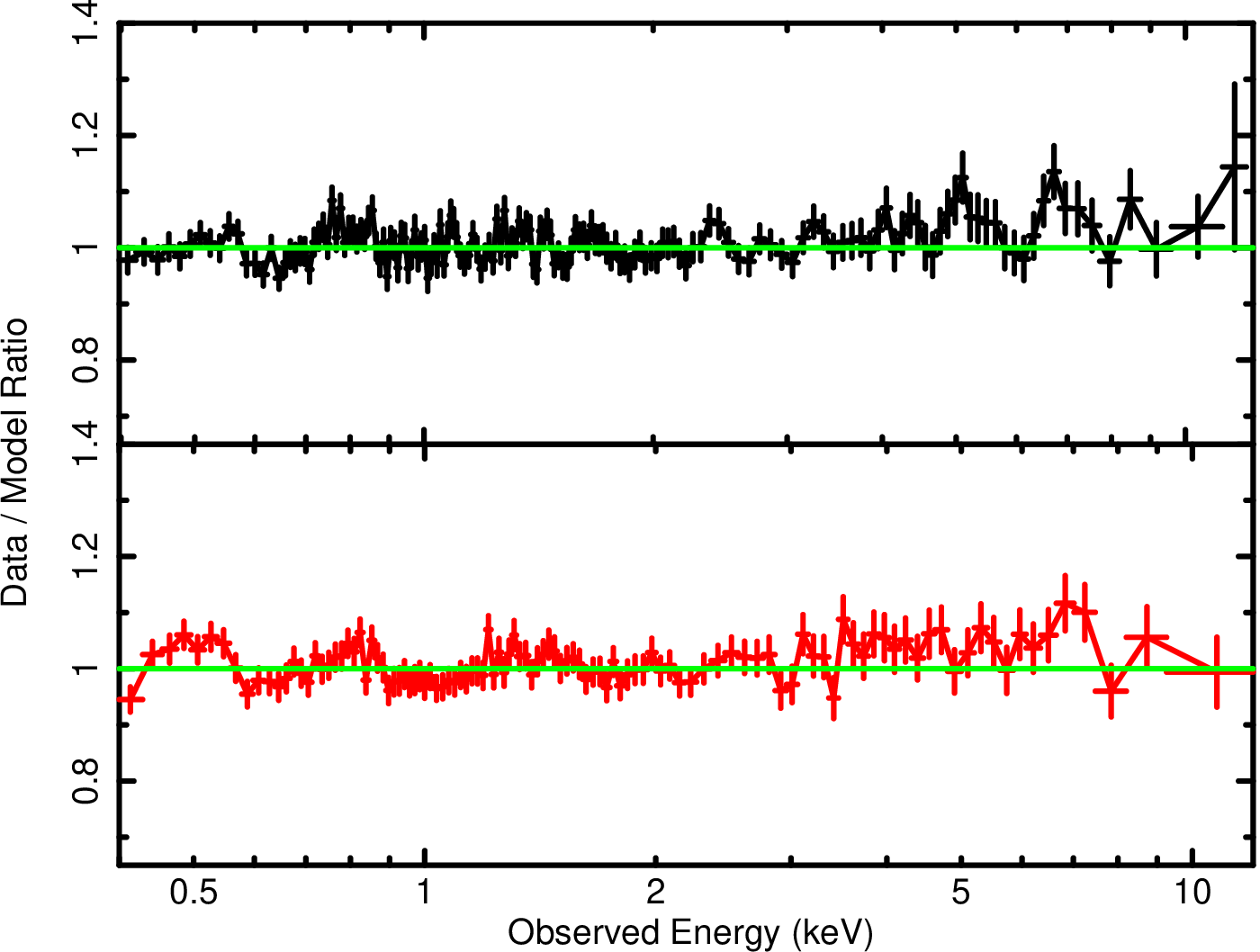}}
\caption{Ratio of \pds\ spectra as observed by the \pn\ camera in Sep 2007 during \xmm\ orbits 1421 (black) and 1422 (red) to a Galactically absorbed ($N_H = 2.0 \times 10^{21}$~cm$^{-2}$) power law ({\it top}), 
with an increased Galactic column density of $N_H = 2.7 \times 10^{21}$~cm$^{-2}$ ({\it middle}), 
and after adding a reflection component 
({\it bottom}).
All spectra are binned to 40$\sigma$ significance, but not more than 20 bins, and are plotted on the same scale to enable easy comparison.
The resulting parameters are summarized in Table~\ref{tab:fit07}.}
\label{xmm07} 
\end{center}
\end{figure}

The September 2007 \xmm\ observations provide the highest S/N spectra of \pds\ to date.
\pn , for example, collected an estimated 437,600 source photons, compared to 121,700 photons during the 2001 observation.
The spectra of both \epic\ cameras and the \rgs\ are of sufficiently high statistical quality to be confronted with a variety of models.
Below 0.4 keV, we find discrepancies of 20\%\ between the \pn\ and \mos\ data with respect to a simple power-law.
Hence, the present modeling is carried out in the 0.4--12.0~keV range and highlights mostly the more sensitive \pn.
These spectra are presented separately for the \xmm\ orbits 1421 and 1422 in Fig.~\ref{xmm07}, along with fits to increasingly complex models.
Although the separation of the data into the two \xmm\ orbits is arbitrary,
it provides an insight into the short-term variability and a reality check for persisting spectral features. 
It is noted that the \pn -fitted models of each orbit also fit the \mos\ spectra very well.
For presentation purposes alone, we have omitted the \mos\ data from the plots.

We first fit the spectra with a simple power law absorbed by the nominal neutral H Galactic column density of $N_H = 2.0 \times 10^{21}$~cm$^{-2}$, using the absorption model of \citet{tbabs}.
We find that the photon slope $\Gamma$ slightly increases by 0.05 between the two orbits.
This model alone is clearly inadequate, as seen by the large residuals plotted in the top panel of Fig.~\ref{xmm07}, and since it yields $\chi ^2$/d.o.f = 1.9.
The residuals below 1~keV suggest the neutral column is underestimated.
Indeed, the best-fit neutral column density is $N_H = (2.67 \pm 0.03) \times 10^{21}$~cm$^{-2}$, which considerably improves the fit to $\chi ^2$/d.o.f = 1.19 (middle panel of Fig.~\ref{xmm07}). 
Separate fits to the \rgs\ and \mos\ spectra yield respective column densities of $(2.64 \pm 0.10) \times 10^{21}$~cm$^{-2}$ and $(2.65 \pm 0.05) \times 10^{21}$~cm$^{-2}$, all consistent and robustly above the Galactic column.
The redshift of \pds\ ($z = 0.184$) allows us to test whether the excess column occurs in the quasar, by invoking a redshifted absorption component in the model.
However, the fit decisively ascribes the entire neutral column to the $z = 0$ component and no column to the redshifted one.
Moreover, if the redshift of the additional component is left to float, the fit requires that it converges to $z = 0$.
The agreement between \pn , \mos , and \rgs\ imply that detector contamination can not be the reason for the excess absorption \citep[see also][]{kirsch05}.
We tested whether perhaps a partially covering absorber can explain the excess low-energy absorption.
However, the best fit model of a partial covering neutral absorber yields a covering fraction of 100\% 
(and with no improvement to $\chi ^2$/d.o.f = 1.19).
There is also no indication of an ionized absorber that could be responsible for the excess absorption, as no absorption lines are detected anywhere in the spectra.
Indeed, the best fit for a partial covering ionized absorber yields a column density that tends to zero, and again no improvement to $\chi ^2$/d.o.f = 1.19.

To make a long story short, the data clearly indicate excess neutral absorption that is consistent with $z~= 0$. Although in principle it is possible that by coincidence this absorption is due to gas in \pds\ that happens to be outflowing by exactly the cosmological recession velocity of the quasar, we reject this notion as highly unlikely.
Moreover, since all indications point to this absorber being neutral, we do not associate it with the highly ionized outflows observed in \pds.
Similar variable Galactic absorption has been detected towards  NRAO 140, where it could be explained by motion in the Perseus molecular cloud complex \citep{turner95}. 
Does the moderately low Galactic latitude of \pds\ of +11$^\circ$ suggest a similar explanation in the present case?
We searched for any molecular clouds towards \pds, both in the IRAS archive, and in the Columbia CO survey \citep{dame87}, but found none.
The origin, therefore, of the excess neutral absorption and its discrepancy with previous measurements remains somewhat unclear.

\begin{figure}[pt!]
\epsscale{0.5}
\begin{center}
\hglue0.1cm
\includegraphics[width=0.57\columnwidth, angle=-90]{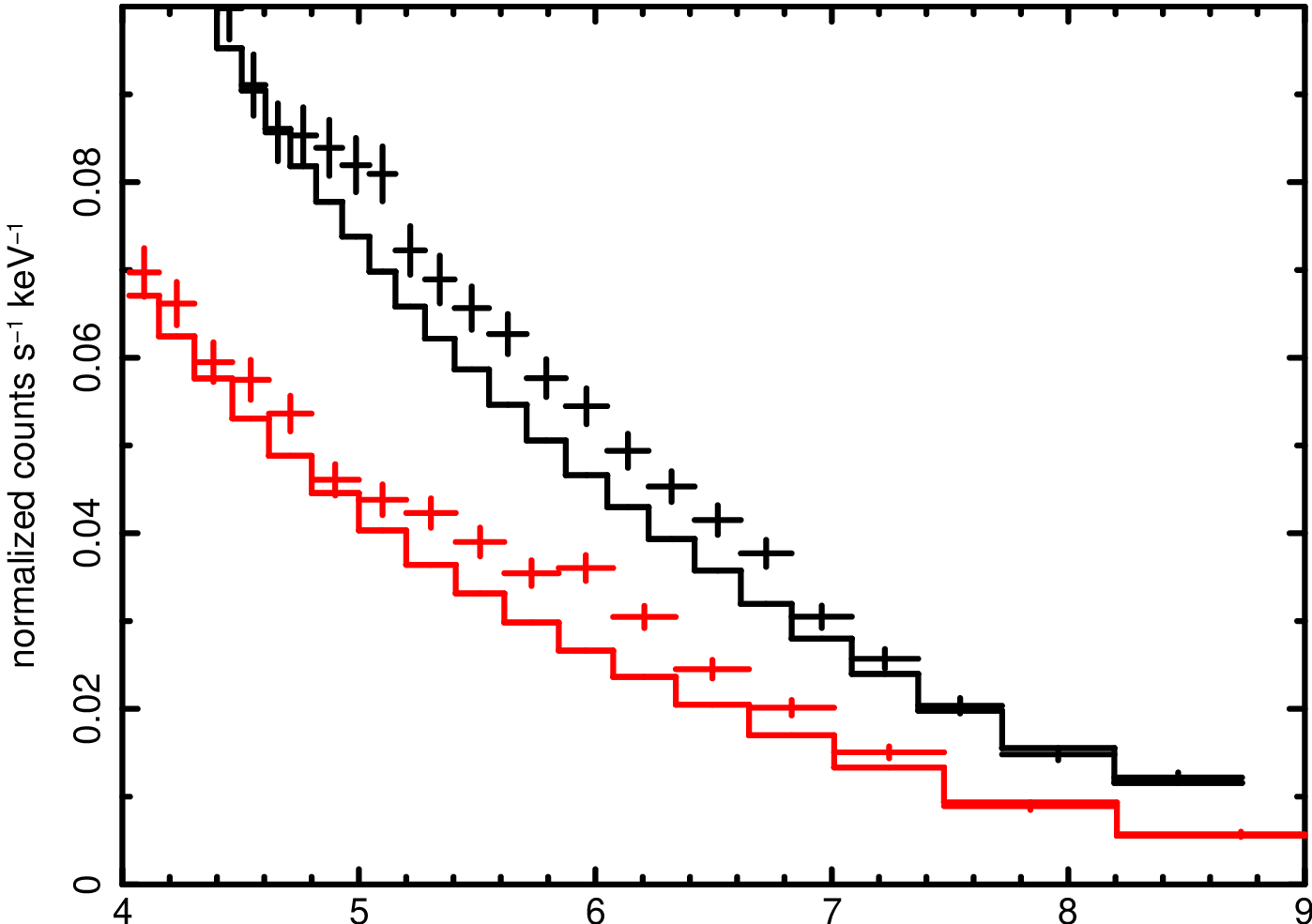} 
\includegraphics[width=0.6\columnwidth, angle=-90]{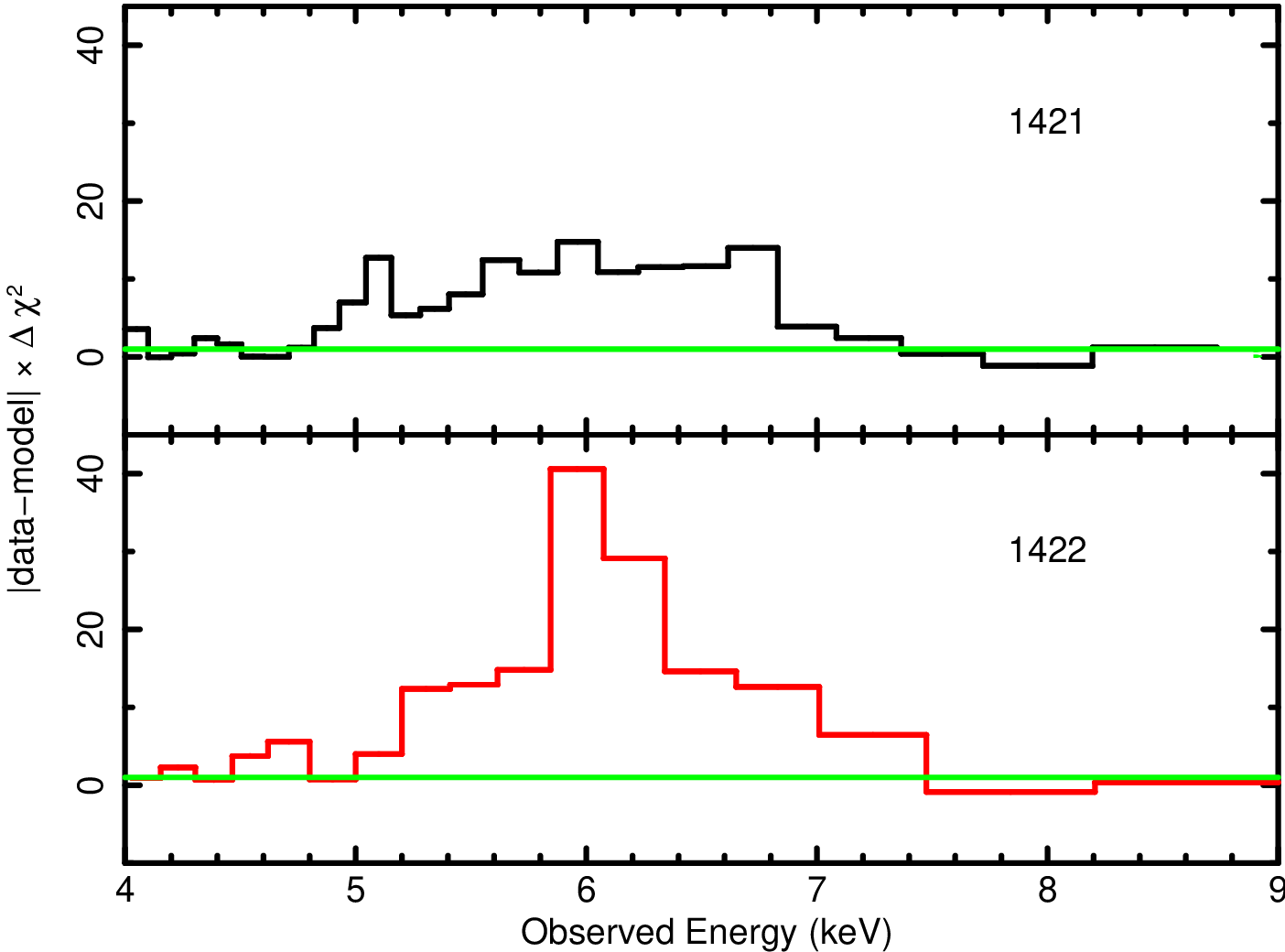} 
\caption{({\it top}) \pn\ spectra and best-fit absorbed power law models (c.f., Fig.~\ref{xmm07} middle panel). ({\it bottom}) Contributions to $\chi ^2$ from \xmm\ orbits 1421 and 1422 around the Fe-K region. Spectra are binned to 40$\sigma$ significance, but not more than 20 bins.
Note the overall similar excess emission from $\sim$~5 to 7~keV and the difference in some spectral details between orbits.
}
\label{FeK}
\end{center}
\end{figure}

In the middle panel of Fig.~\ref{xmm07}, we show the spectral residuals that remain after incorporating the higher column density in the model.
It can be seen that these residuals are fairly similar in both orbits (but keep in mind the slightly different power law slopes).
The residuals below 1~keV could be due to the incomplete absorption model around the O-K ($\sim 0.53$~keV) and Fe-L ($\sim 0.71$~keV) photoelectric edges.
At high energies, there is excess emission around the Fe-K region between 5 and 7 keV in the observed frame.
Note that much of this emission is blueshifted with respect to the quasar frame, given that the 6.4 -- 7.0~keV range of Fe-K lines corresponds in the observed frame (Fig.~\ref{xmm07}) to the 5.4 -- 5.9~keV energy range.
The Fe-K excess emission and its high statistical significance is further demonstrated in the $\chi ^2$ residuals plotted in Fig.~\ref{FeK}.
The second data set (orbit 1422) has a narrow component, but both data sets feature large broad residuals resembling the electron-scattered blue wings typical of ionized (Compton) reflection.
Also important is the over all curvature of the model between 1 and 5 keV, where no narrow features occur, that is not properly reproduced by an absorbed power law.
Together with the Fe-K residuals this strongly suggests a reflection spectral component.
Indeed, adding a constant density reflection component \citep[{\it reflion}, by][]{ross05} notably and significantly improves the agreement of the model with the data ($\Delta \chi ^2 = 313$, and $\chi ^2$/d.o.f = 1.05).
The broad-band spectral curvature between 1 and 5 keV is also much improved when reflection is included in the model, as can be seen by comparing the middle and bottom panels of Fig.~\ref{xmm07}. 
The small remaining residuals are of order of 10~-- 15~\%.
The best-fit parameter values for the two combined orbits are given in Table~\ref{tab:fit07}.
When fitting the two data sets separately, we find that the reflection component fits the second data set (orbit 1422) particularly well, yielding $\chi ^2$/d.o.f = 0.99. The fit to the first spectrum (orbit 1421) is somewhat poorer with $\chi ^2$/d.o.f = 1.10.  Models fitted separately to each spectrum yield parameters that are similar to those of Table~\ref{tab:fit07} ($\log \xi \approx 4, A_{\mathrm Fe} \approx 3, z \approx 0.1$), but are not as  tightly constrained. 

We tested alternative models to reflection, including a plain Gaussian emission line and a gravitationally distorted emission line.
Both give unphysical results as could have been expected from the spectral shape of the residuals and the combined narrow and broad nature of the residuals in the middle panel of Fig.~\ref{xmm07}.
A Gaussian fit forces an enormously broad feature with $\sigma$~= 3~keV, and centered at an unreasonably low energy of $\approx 4$~keV in the quasar frame, essentially mimicking the underlying broad reflection continuum.
A relativistically distorted line profile \citep{laor91} also yields unreasonable physical parameters, such as a disk inclination of almost 90$^\circ$ and an emissivity that {\it increases} with radius.
Moreover, partial covering neutral, and partial covering ionized absorption models were also tried in place of reflection, but could not eliminate the residuals observed in the middle panel of Fig.~\ref{xmm07}.
In summary, we prefer the reflection model, which is most appropriate for describing both the broad-band and the Fe-K spectrum of \pds\ during the 2007 \xmm\ observations.

\begin{figure}[pt!]
\begin{center}
\includegraphics[width=0.75\columnwidth, angle=-90]{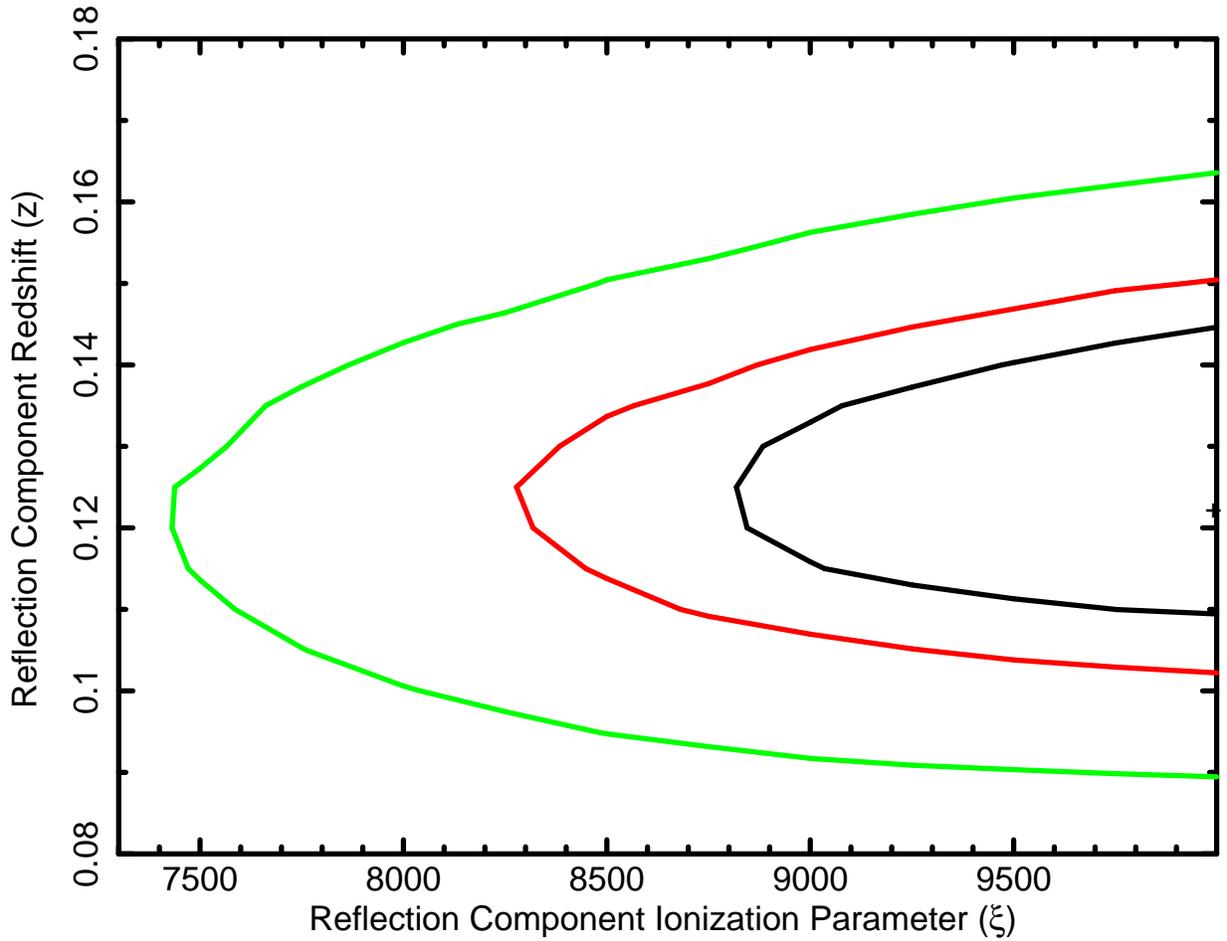} 
\caption{Two-parameter statistical confidence contours of the ionization parameter ($\xi$) and redshift ($z$) of the reflection component fitted to the 2007 \pn\ spectra of \pds.
The three contours represent $\Delta \chi ^2$ of 2.3, 4.61 and 9.21 with respect to the best values of $\xi$ = 10000 and $z$ = 0.122.
Values of $\xi > 10000$ are not available with the current reflection model, but the high confidence in the result of $z < 0.184$ (i.e. outflowing reflector) appears to be robust.
}
\label{contours}
\end{center}
\end{figure}

The fit of the reflection model to the combined 2007 spectrum constrains the frame of reference of the reflection component rather tightly.
This comes from the position of the reflected Fe line (for $\log \xi \approx 4$ primarily H-like at 6.97~keV) and even more from the electron-scattered blue wings of the line that give it a slightly broadened and skewed profile \cite[see, e.g., Fig.~3 in][and also Fig.~\ref{abs_models} below]{ross05}.
Interestingly, the best-fit redshift for the reflection model is indicative of a frame of reference that is distinct from both the quasar and the observer, namely $z = 0.12 \pm 0.02$ to be compared with $z = 0.184$ for \pds.
Fixing the reflection redshift to that of the quasar and refitting results in an inferior fit ($\Delta \chi ^2 = 28$ resulting in $\chi ^2$/d.o.f. = 1.07).
The intermediate reference frame can be understood as reflection off of fast moving gas,
perhaps the outflows detected in previous observations? 
Since both the frame of reference and the ionization state of the reflecting medium bear on the position of its Fe-K line, it is important to investigate the connection between the two in the fit.
In Fig.~\ref{contours}, we plot the combined statistical confidence contours for the redshift and the ionization parameter. 
The plot demonstrates that the kinematical shift (i.e., outflow) of the reflector is statistically robust ($> 3\sigma$) against the ionization uncertainties within the framework of the present model.
However, this result is model dependent and a realistic outflow is surely more complicated than the slab reflection used here.
The regime of ionization parameter beyond $\xi~=$ 10,000 (cgs units) is inaccessible with the current reflection models, and therefore is not plotted in Fig.~\ref{contours}.
Moreover, the constraints on each \xmm\ data set when fitted separately are insufficient to produce similar contour plots, only the combined fit is well constrained.
On the other hand, much higher ionization ($\xi~>$ 10,000 in cgs units) can not shift the line position to beyond its current H-like position, which suggests that even higher ionization would perhaps not invalidate the outflowing nature of the reflector.
There is also no obvious indication by extrapolation of Fig.~\ref{contours} that higher ionization compromises the redshift.
The putative radial velocity of the reflector, which would be moving more or less parallel to
but outside the line of sight is
$v_\mathrm{out} / c \approx (0.12 \pm 0.02) - 0.184 = -0.06 \pm 0.02$, 
or $v_\mathrm{out} = -18,000~\pm 6,000$~\kms.
In the next section, we show that the moving reflection component may be a persistent feature of the otherwise variable \x\ source in \pds.

Furthermore, if the observed reflection component is due to reflection of the observed primary power law,
it is possible to estimate the solid angle occupied by the reflector as viewed from the primary source.
In a simplified geometry with perfect reflection (i.e., no flux losses on the reflector), and neglecting time lag, the flux ratio of the reflection component to the primary power law gives a rough idea of this fractional sky coverage $\Omega _\mathrm{refl}/ 2\pi$.
The no-losses approximation is reasonable for electron reflection in a nearly fully ionized reflector.
The \citet{ross05} model assumes uniform illumination of the reflector and a plane parallel geometry.
It computes the reflection integrated over all angles from 1~eV to 1~MeV by a power law source ranging from 1~eV to 0.3~MeV.
For the present models, the flux ratio in the observed region from 0.4 -- 12.0 keV gives essentially the same result as that in the full range.
For the best fit model, this flux ratio implies $\Omega _\mathrm{refl}/ 2\pi \approx 0.2$.
The best-fit reflected flux in the present models happens to be similar in both orbits, despite the decrease by approximately 30\% of the mean primary power law flux.
This indicates the reflector is likely extended from the source.
Given the multi-component nature of the \pds\ \x\ spectrum, and given the variability of shifted emission and absorption on short and long time scales, as well as the low Galactic latitude (+11$^\circ$) of \pds , it is worthwhile to examine the viability of a previously unknown Galactic source along the line of sight.
An \x\ binary (XRB) could be the primary candidate for such a source.
In order to test such a conjecture, we inspected the \x\ \chandra\ image, the \uv\ \xmm\ optical monitor (OM) image, and the 2MASS K-band image of \pds. 
They are all perfectly consistent with point sources and are co-aligned to within a fraction of an arcsecond.
By extrapolating the $\log N - \log S$ distribution of \citet{grimm02} for galactic XRBs down to a flux, say, of $f = 10^{-12}$ \cgsflux\ \citep[about 10\% of the current flux of \pds, but two orders of magnitude fainter than the completeness limit of][]{grimm02}, one can use the expression $N(>S) = 72(f/3.2\times 10^{-10}$ \cgsflux )$^{-0.41}$ to estimate the number of XRBs in the sky that are brighter than 10$^{-12}$ \cgsflux\ to be not more than 800.
Consequently, the probability for a random XRB within the solid angle of 0.5\arcsec $^2$ is $< 4\times 10^{-10}$; negligibly small.  In other words, it seems unlikely that a foreground XRB is present along the line of sight to \pds.

In Section~\ref{sec:partabs}, we explore partial covering models by outflows as a way to explain on an equal footing the entire slew of \x\ spectra obtained for \pds\ over the past decade.
In the 2007 \pn\ spectra shown in Fig.~\ref{xmm07} there is no obvious absorption by an outflow, and thus the absorption models actually do not improve the fit. 
The reflection component, on the other hand, seems to persist in all epochs.

\subsection{Outflow?}
\begin{figure}[pt!]
\begin{center}
\includegraphics[width=0.75\columnwidth, angle=-90]{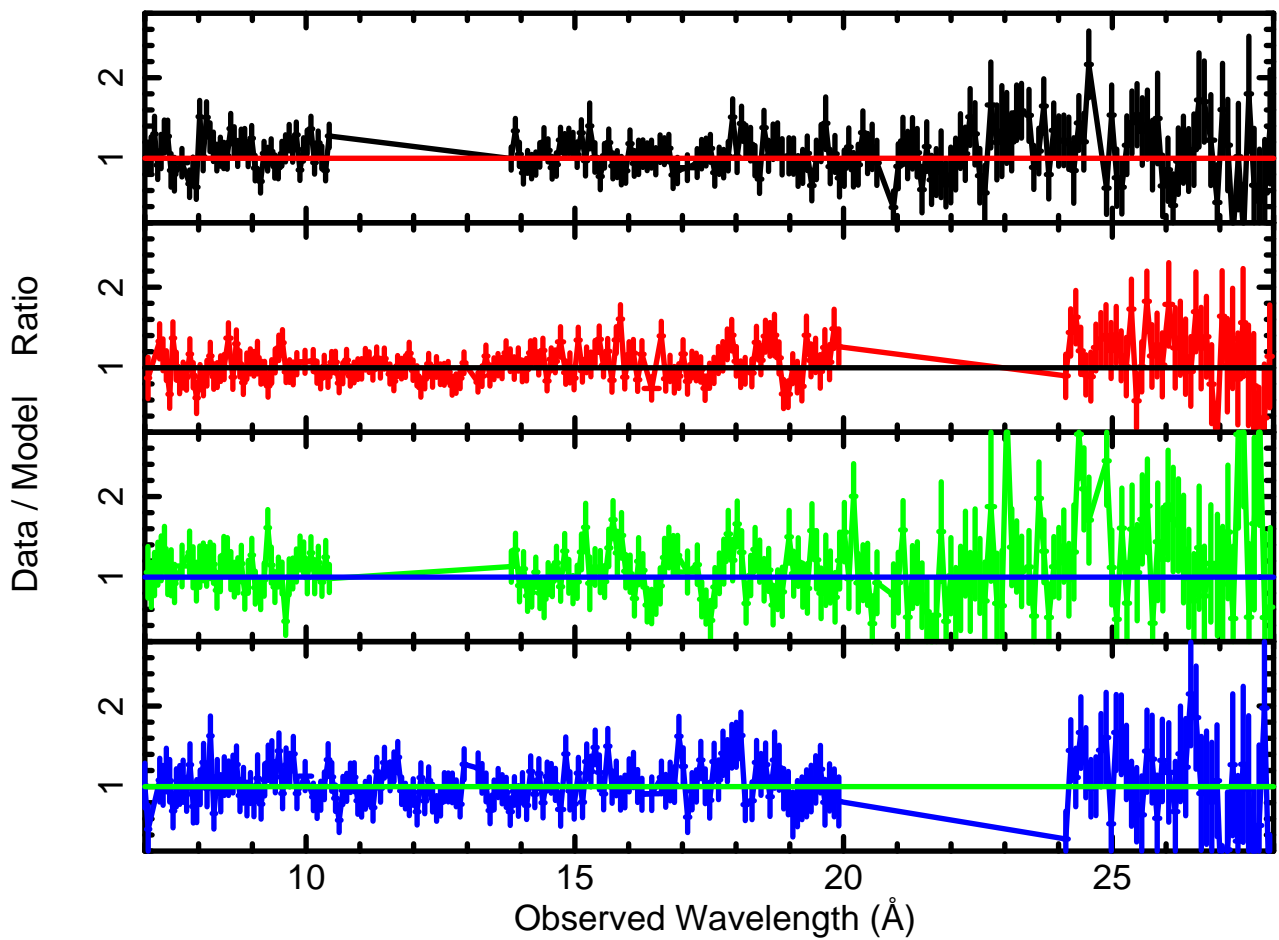} 
\caption{ 
Ratio of \pds\ \rgs\ spectra to the best (continuum) model obtained from fitting the \pn\ spectra.
Data are shown separately from top to bottom for \xmm\ orbit 1421: \rgs 1, \rgs 2, and for orbit 1422: \rgs 1, and \rgs 2, and are rebinned uniformly by a factor of two.
Straight line segments with no errors are due to gaps in the \rgs\ detectors.
No absorption (i.e., outflow) features recur consistently in several data sets.}
\label{RGSpanels07}
\end{center}
\end{figure}

\begin{figure}[pt!]
\begin{center}
\includegraphics[width=0.75\columnwidth, angle=-90]{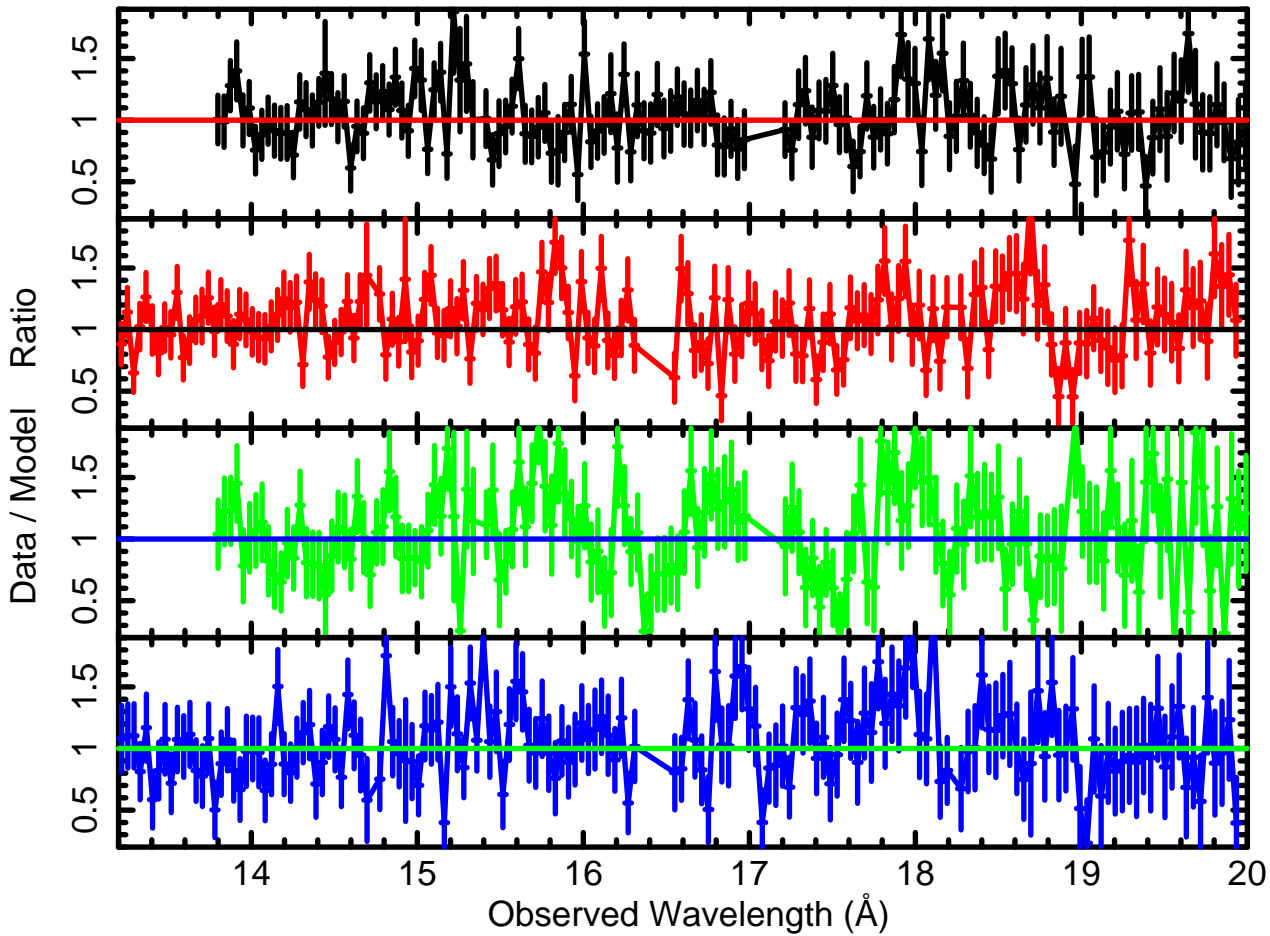} 
\caption{
Zoom into ratio of \pds\ \rgs\ spectra to the best (continuum) model obtained from fitting the \pn\ spectra.
Data are shown separately from top to bottom for \xmm\ orbit 1421: \rgs 1, \rgs 2, and for orbit 1422: \rgs 1, and \rgs 2, and are not rebinned beyond the original reduction (Section~\ref{datareduction}).
Straight line segments with no errors are due to gaps in the \rgs\ detectors.
No compelling evidence for recurring absorption (i.e., outflow) is seen.}
\label{RGSzoom07}
\end{center}
\end{figure}

We searched the present 2007 data set for the fast outflows detected by \citet{reeves03, reeves09}.
The high-energy region of the \pn\ spectrum during orbit 1421 is shown in Fig.~\ref{FeK}.
No sign of absorption can be detected whatsoever;
Neither the lines that appear in the \suz\ spectrum at 7.67~keV and 8.13~keV \citep{reeves09}, nor the edge that appeared in the 2001 \xmm\ observation \citep{reeves03} are seen in this data set. 
We further inspected the \rgs\ spectra from both \xmm\ orbits.
Fig.~\ref{RGSpanels07} presents the ratio of the \rgs\ spectra to the model used above to fit the \pn\ spectra, with no adjustments whatsoever.
Since this model includes no discrete spectroscopic features (except perhaps the neutral O absorption at $\approx$ 23.5~\AA), we would expect absorption (or emission) lines or edges due to the outflow to show up in the \rgs\ data. 
We show in Fig.~\ref{RGSpanels07} each \rgs\ (1 and 2) and the data of each orbit separately in an attempt to identify outstanding features. 
The statistics of the data presented is rather high, ranging from over 5,000 to over 10,000 source photons in each panel, which are distributed over $\approx$~170 bins.
In fact, this is the best high-resolution \x\ spectrum of \pds\ to date.
A closer look into the 13--20~\AA\ spectral region, where the \rgs\ is most sensitive, is presented in Fig.~\ref{RGSzoom07}.
We are unable to identify any recurring lines that appear in all of the panels of Figs.~\ref{RGSpanels07} or \ref{RGSzoom07}. 
At the very least, we would expect genuine absorption lines to be present in both RGSs during a single observation (panels 1 and 2, or 3 and 4), but we find none.
We conclude that there is no unambiguous sign in September 2007 of the outflows detected in the two previous observations.
The intermittent nature of the outflow has immediate implications on the overall AGN energy budget as affected by such flows.

\section{Spectral Variability}
\label{sec:partabs}

\begin{figure}[pt!]
\begin{center}
\includegraphics[width=0.75\columnwidth, angle=-90]{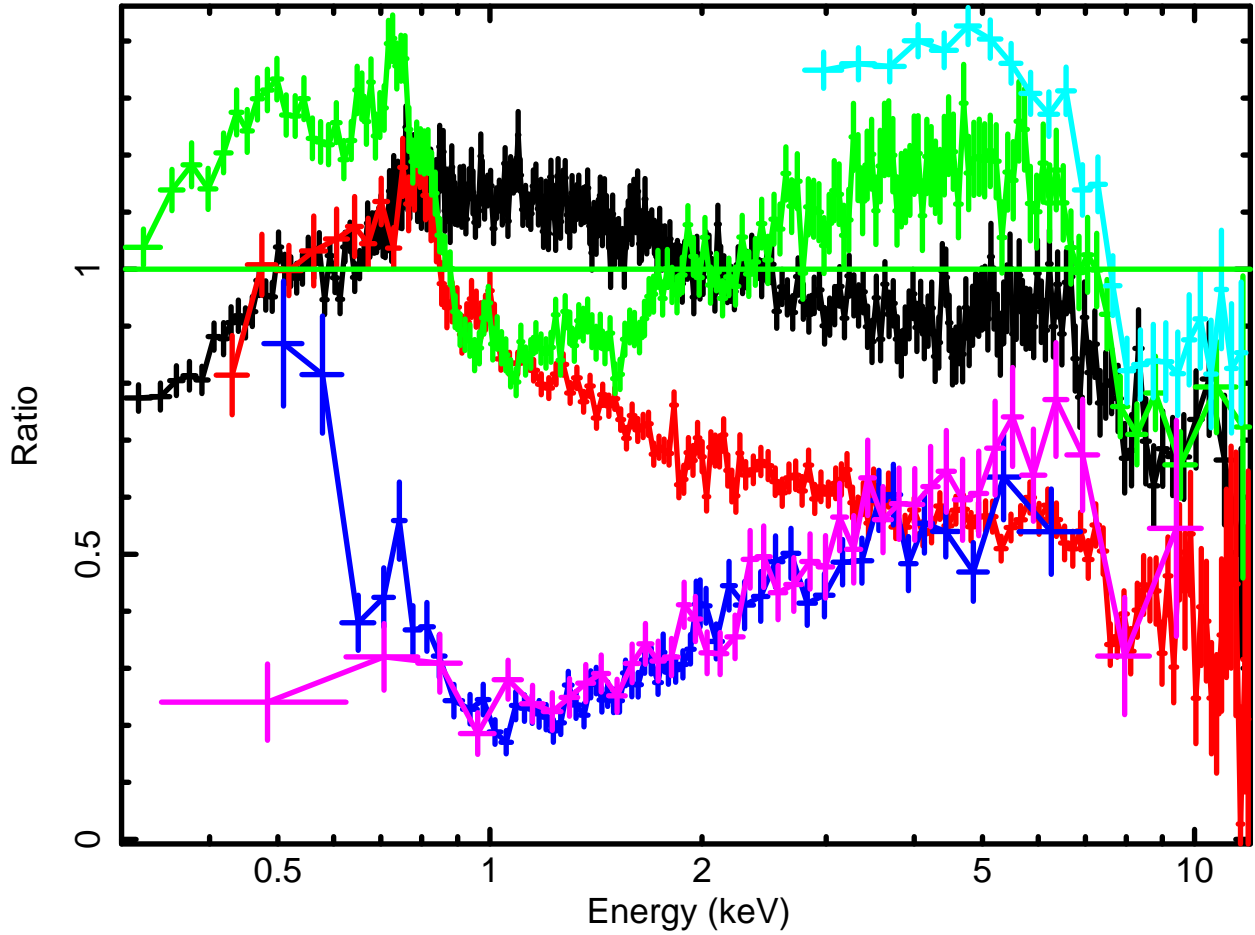} 
\caption{\label{fig:comp} PDS 456 as observed by \pn\ in Sep. 2007 (black), Suzaku in 2007 (red), \pn\ in 2001 (green), Chandra/MEG in 2003 (blue), ASCA/SIS in 1998 (magenta), and RXTE/PCA in 2001 (cyan). The ratio of data to a Galactically absorbed power-law spectrum with photon index $\Gamma = 2$ is plotted for all observations.
}
\label{fullcomp}
\end{center}
\end{figure}

The appearance and disappearance of the outflow in \pds\ within several months is only one aspect of the full range of changes taking place in this intriguing quasar.
Since it was discovered, \pds\ was observed by almost all available \x\ telescopes.
Therefore, one can expect to obtain a more comprehensive picture of this source by studying its \x\ spectral history.
In addition to the current data sets from the two \xmm\ orbits in 2007,
there are five archival \x\ observations of \pds.
These are available from \xmm\ (February 2001), \suz\ (February 2007), \chandra\ (May 2003), \asca\ (March 1998), and \xte\ (February-March 2001).
With the exception of \chandra, all observations were reported in the literature
 \citep{reeves00, reeves03, reeves09}.
For the details about these data and how they were reduced we refer the readers to those references.
The \chandra\ \hetg\ observation was not published because it caught \pds\ in a very low state.
The data are reduced here using CIAO version 3.2.1 and CALDB 3.0.1 and the standard CIAO threads.
Spectral fits were carried out simultaneously for both \hetg\ instruments (MEG and HEG), 
although we show only the MEG spectra.
In Fig.~\ref{fullcomp}, we plot the ratio of the different \pds\ spectra to a simple Galactically absorbed ($N_H = 2 \times 10^{21}$~\cmii) power-law with photon index $\Gamma = 2$.
From the 2007 \xmm\ campaign, only the spectrum of orbit 1421 is presented, to minimize crowding of the plot. 
The total flux variability is of the order of a factor of 2--3.
This may not be expected from a quasar as bright as \pds\ ($L_X \approx 10^{47}$~erg~s$^{-1}$).
However, \pds\ has already been observed varying by a factor of a few over less than a day \citep{reeves00}.

From Fig.~~\ref{fullcomp}, it can be seen that spectral structure below 1~keV varies both overall and in its narrow features.
The Fe-K region beyond 5~keV also changes drastically.
What appears as an absorption edge, and can be interpreted as an outflow, varies both in position and in depth.
The significance of emission features, possibly identified as due to reflection, changes as well.
Between 1~keV and 5~keV, the spectral curvature changes in what seems to be an intricate combination of emission and absorption variability.
This observed complexity can not be characterized by a single spectral component or one obvious spectral parameter, such as power-law slope, Fe-K line flux, or absorbing column density.
What on the face of it seem to be common features in several spectra,
such as an Fe-K absorption edge (or emission line?) beyond 6~keV, 
or an Fe-L emission hump below 1~keV, 
upon closer inspection in fact appear to occur at slightly different energies in each spectrum.

In the next section, we attempt to use one possible type of models to fit all of the spectra of \pds.
For this exercise we fit separately six epochs with high quality CCD data at an energy resolution of $\Delta E \approx 100$~eV, or better, namely those of \pn\ (two \xmm\ orbits in September 2007), \pn\ (2001), \suz\ XIS 0 \& 3 (February 2007), \chandra\ \hetg\  (2003, MEG \& HEG) rebinned to CCD resolution due to low S/N, and \asca\ SIS (1998). 
We leave out the \xte\ spectrum due to its lower resolution.
Data from \mos\ on \xmm\ and XIS~1 on \suz\ are less sensitive at high energies, but their data are consistent with those of \pn\ and XIS 0 \& 3, respectively \citep[see][]{reeves03, reeves09}.
The six data sets used here are listed in Table~\ref{obs_log}. The modeling strategy is described in the next section.

\subsection{Partial Covering Approach}
\label{sec:pc}
For the model, we invoke a simple continuum power law source, partially covered by a (possibly ionized) absorber and absorbed locally in our Galaxy.
For the absorbing outflow, the analytical {\it warmabs} models version 2.1ln8 of \xstar\ \citep{xstar01} available in \xspec\  are used \footnote{http://heasarc.gsfc.nasa.gov/docs/xanadu/xspec/models/xstar.html}.
A reflection component is also included as in Section~\ref{sec:xmm07}, to account simultaneously for Fe-K lines and for the soft excess that is observed in some data sets and apparently varies in its discrete features (Fig.~\ref{fullcomp}).
In \xspec\ code, the model used is {\it TBabs} ({\it zpowerlw + zpowerlw * warmabs * warmabs + reflion}).
That partial covering of the central source can provide useful insights into the AGN geometry was realized by \citet{reichert86}.
These types of models provide a wide range of broad band spectral curvature that may (or may not) be able to produce the vastly different spectra seen in Fig.~\ref{fullcomp}. 
The model discussed above for the 2007 \xmm\ spectra (Section~\ref{sec:xmm07}) where no absorption is detected qualifies as the degenerate extreme case of zero coverage of the source.
Realistic partial covering models also include narrow absorption lines, which impose additional constraints complementary to the broad-band curvature of the spectrum.
However, when $\chi ^2$ minimization techniques are used to select models, as in the present work,
the continuum statistics overwhelms that of the discrete features.
Thus, the narrow features are much harder to constrain with the current CCD spectra.
Similar models of partial covering by an ionized outflow have been tried with relative success by \citet{reeves03} for the 2001 observation of \pds, and recently by \citet{miller08} for \mcg.

The modest goal of this section is to test the conjecture that perhaps it is the varying fractional coverage of the source by the outflow that drives the spectral variability, and whether these models can provide an acceptable overall explanation to {\it all} of the six spectra of \pds.
This type of partial coverage models is complex and invokes by far more parameters than can be constrained by the data.
Consequently, it is beneficial to fix a-priori as many parameters as possible.
The illuminating spectrum to be absorbed and reflected is the aforementioned power law with its slope fixed to $\Gamma = 2.25$, which is the best fit value for the high S/N and relatively featureless spectra obtained with \pn\ (orbit 1422) and with  \suz.
Two ionization components are included for the absorber.
These components represent high and low ionization in an absorber that more realistically could have a continuous distribution of charge states \citep{steenbrugge03, behar09}, and are therefore not well constrained by low-resolution data.
The fractional coverage of the power law source by the outflow is a fitted parameter.
The outflow velocity is fitted in each spectrum and is found to somewhat vary from one observation to the other. The velocity broadening is fixed at $v_\mathrm{turb}$ = 2,500~\kms, which fits well the only grating measurement of absorption lines in \pds\ (see Section~\ref{sec:rgs01} below) and is consistent with the width of the absorption lines in the \suz\ spectrum \citep{reeves09}.

Since the reflection component in the 2007 \xmm\ spectra appeared to be moving (Section~\ref{sec:xmm07}) and inconsistent with the rest frame of \pds,
it may be moving with the absorbing medium.
Consequently, we tie the outflow velocities of the reflection and absorption components,
which would be a good approximation as long as the opening angle of the putative reflecting outflow is not too large.
Abundances are set to their solar values, except for the Fe abundance of the reflector that is set at three times solar as required by the fits in Section~\ref{sec:xmm07}. 
No independent Fe-K absorption or emission feature beyond the outflow and reflection features is added.
Also for simplicity, no relativistic blurring is invoked, since it tends to 
make the models even more degenerate than they already are.
Despite all of these impositions on the model, it is still fairly degenerate and the fit can not reliably constrain all of its free parameters.
To that end, the \suz\ spectrum can be singled out with its detected absorption lines from two kinematic components and the ionization parameter that can be more accurately determined \citep{reeves09}.
Consequently, in our model for the \suz\ spectrum, the absorber ($v/c \approx -0.3$) and reflector ($v/c \approx -0.12$) are allowed to have different velocities. 


\subsection{Fit Results}
\label{fit_results}

\begin{figure}
\begin{center}
\vglue0.0cm
{\includegraphics[angle=-90,width=8cm]{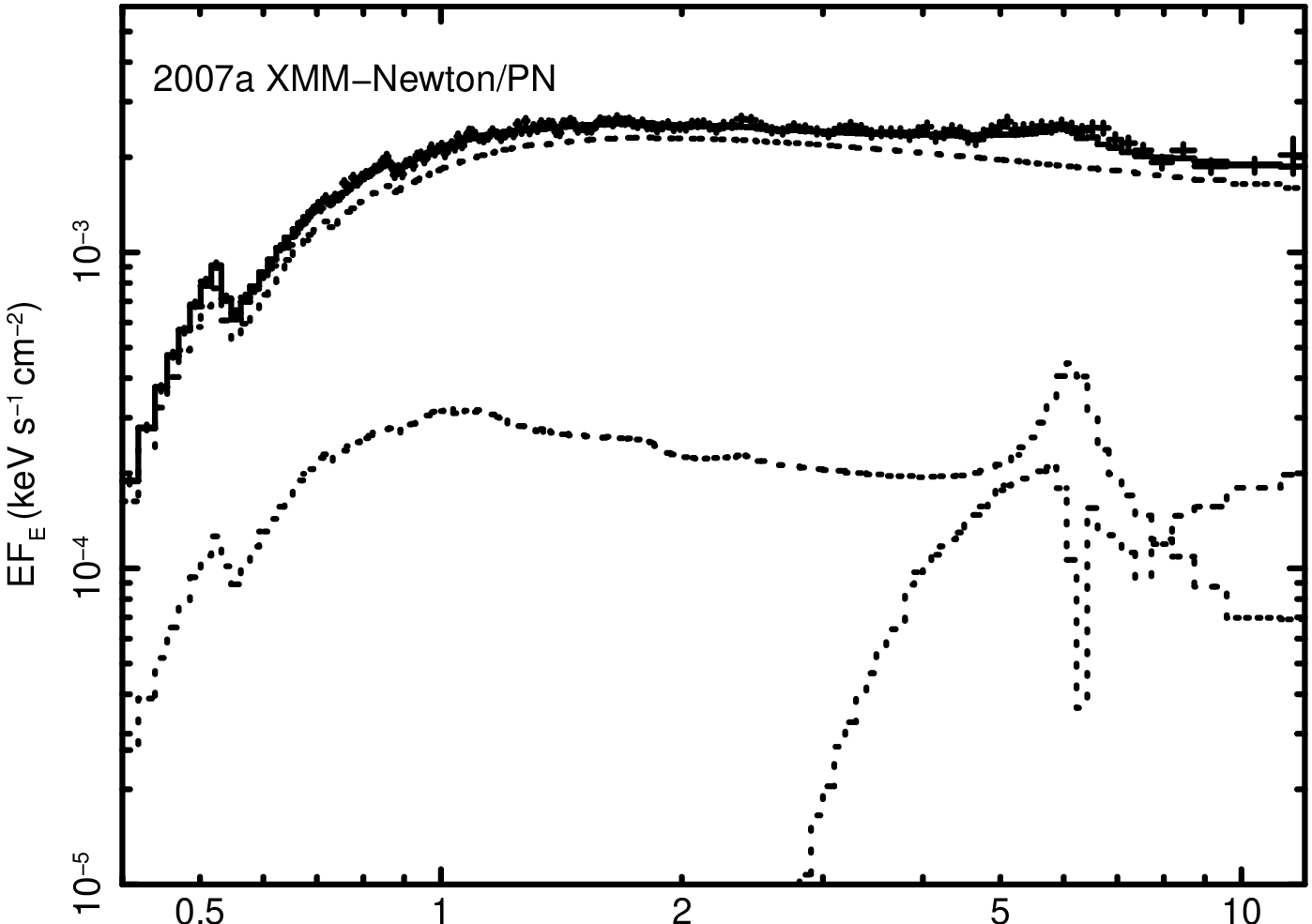}}
{\includegraphics[angle=-90,width=8cm]{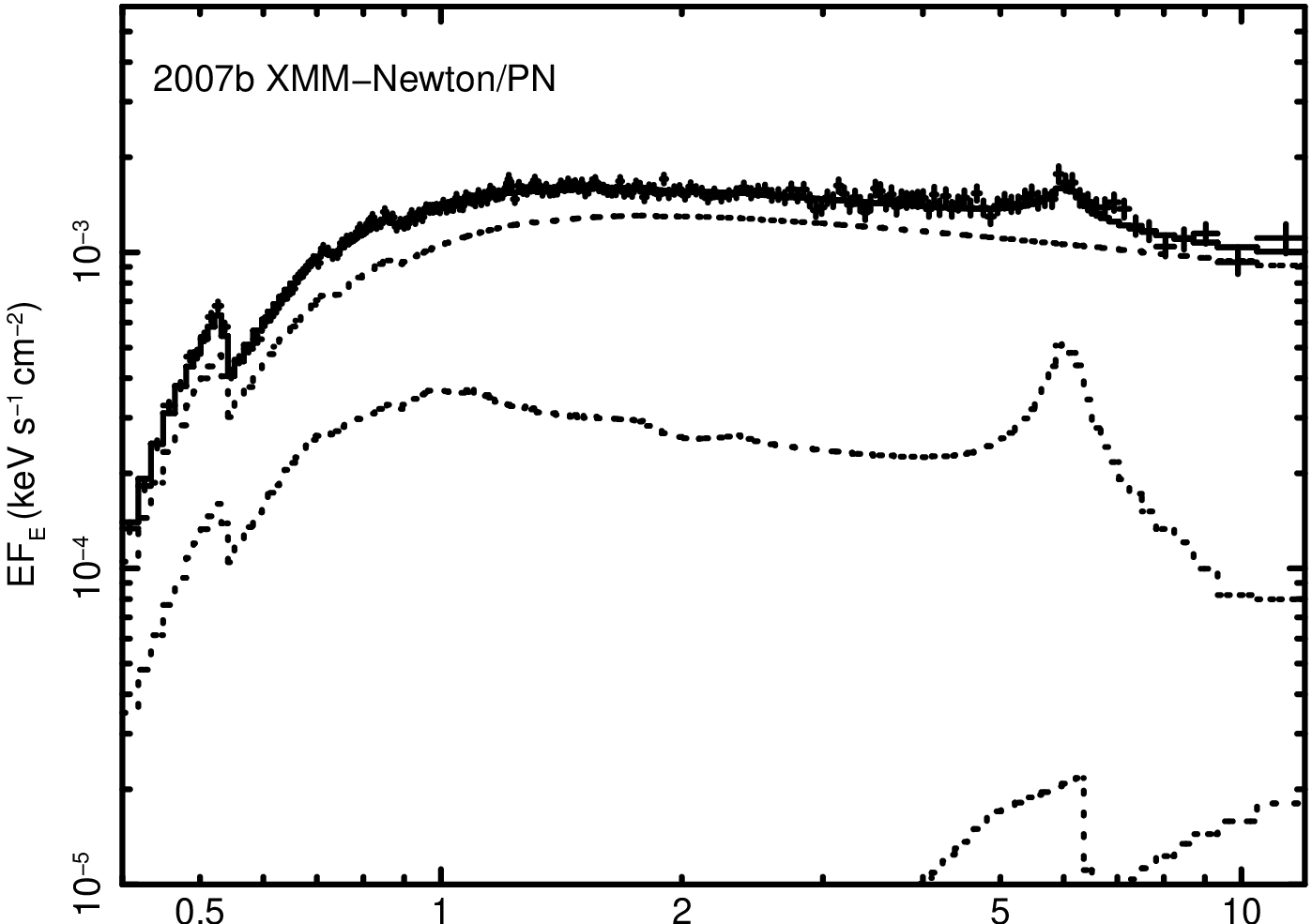}}
{\includegraphics[angle=-90,width=8cm]{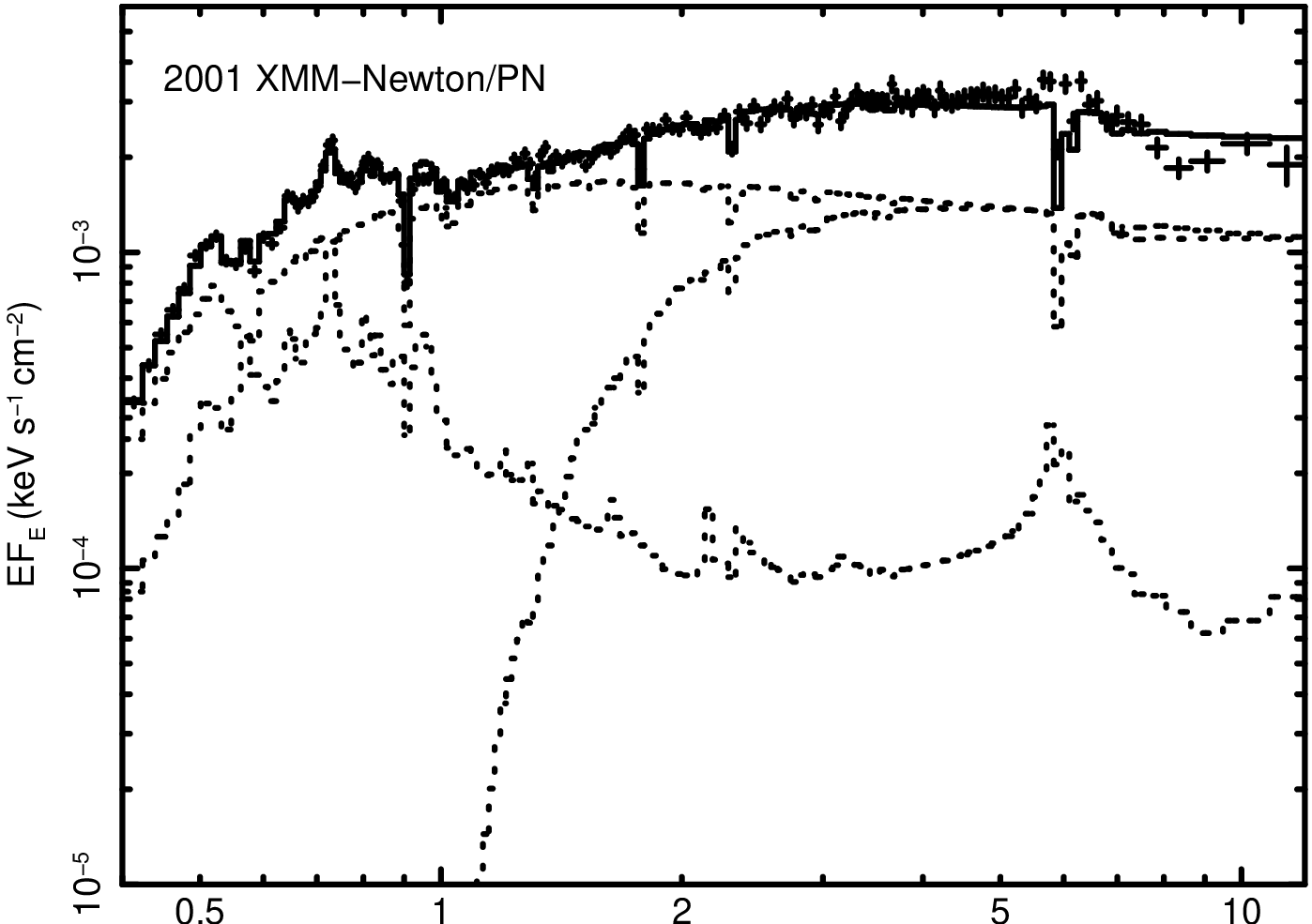}}
{\includegraphics[angle=-90,width=8cm]{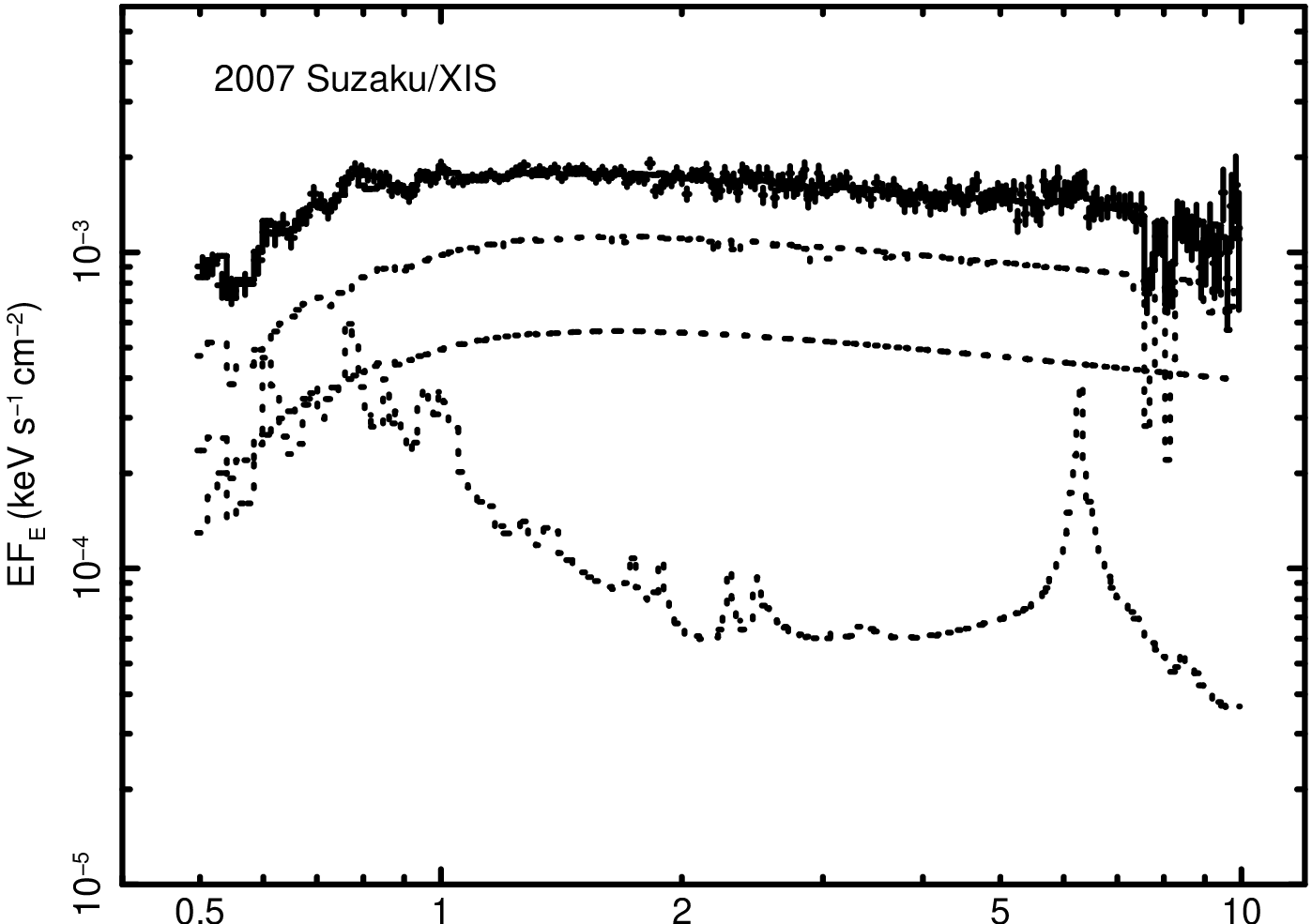}}
{\includegraphics[angle=-90,width=8cm]{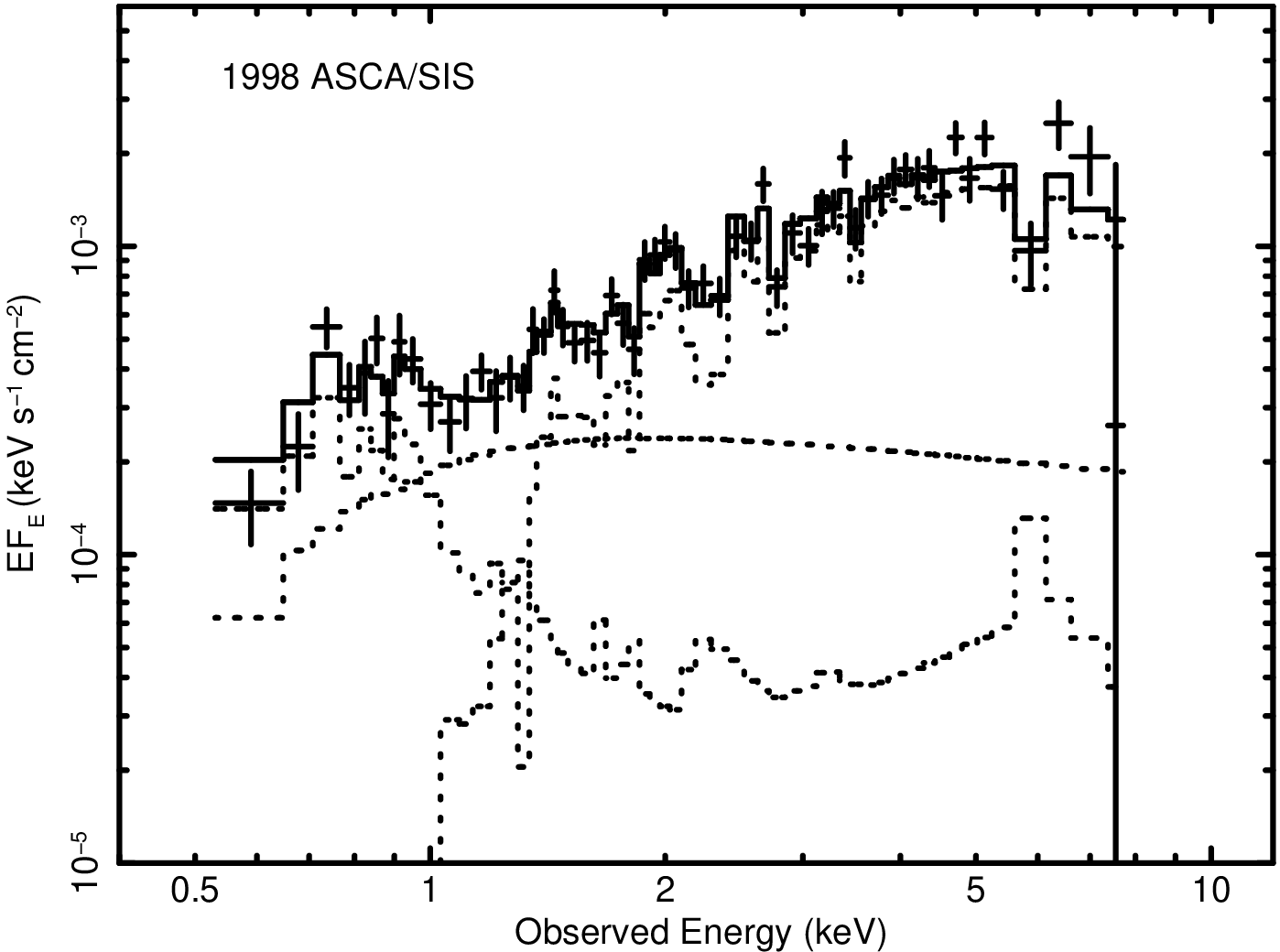}}
{\includegraphics[angle=-90,width=8cm]{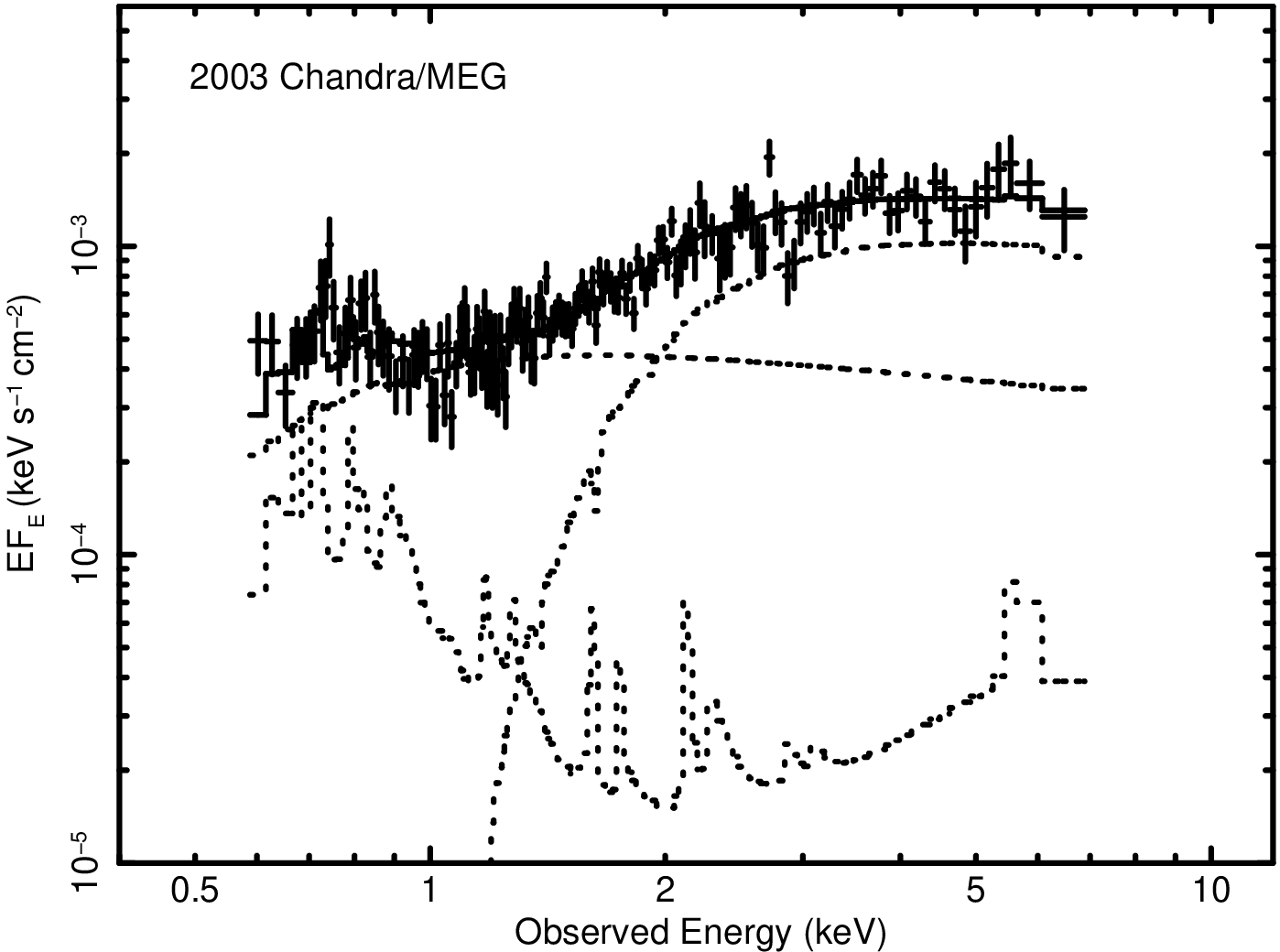}}
\caption{Unfolded spectra of \pds\ sampling its spectral variability over the past decade (in arbitrary order).
Spectra are produced using the partial covering model discussed in the text.
The three dotted lines in each spectrum delineate the absorbed, and unabsorbed power law components, and the reflection component.
The resulting parameters are summarized in Table~\ref{table:fit}.
}
\label{abs_models} 
\end{center}
\end{figure}

Using the partial covering model with its restricted parameters as described above, 
we fitted the six available spectra.
The results are presented in Fig.~\ref{abs_models} in the form of unfolded spectra.
It is worth warning that unfolded spectra should not be used to confirm spectral features in the data, but rather to present what a best-fit model spectrum would look like assuming it well represents the source spectrum.
In each panel of Fig.~\ref{abs_models}, the different model components are plotted, namely the absorbed power law that turns over above 2~keV, the unabsorbed power law that turns over only below 1~keV due to Galactic  absorption, and the reflection component.
The relative contributions of the absorbed and unabsorbed components, i.e. the partial covering, varies dramatically between epochs. 
The absorbed component is very weak in the 2007 \xmm\ observations (top panels of Fig.~\ref{abs_models}), as the fit provides only upper limits to its presence and indicates very low covering fraction ($<20\%$), if any.
Such an absorber would be consistent with the highly absorbed hard \x\ component detected with \suz\ by \citet{reeves09}, but we can not conclusively confirm its presence in the 2007 \xmm\ spectra.
The reflection contribution to the total flux is generally small, but persistent in all epochs, and its 
emission below 1~keV is required to produce the soft excess observed in some of the spectra as well as the Fe-K line emission and the spectral drop off above 5~keV. 
It can be seen from Fig.~\ref{abs_models} that the models provide reasonable representations of the wide variety of spectra observed from \pds.
Despite the several fixed parameters and most notably the power law slope,
by varying the column densities and the fractional covering by the outflow,
the quality of the fits is reasonable with $\chi ^2$ / d.o.f $\approx$ 1.0 -- 1.3, although not always statistically acceptable. 
We believe the imperfect fits are not necessarily due to the inadequacy of partial covering for describing \pds\ at all epochs.
More likely, it is the shortcomings of the models and their incomplete detail.
In some cases, calibration issues can also affect the fits.
For example, ignoring the instrumental Si edge region of the \suz\ spectrum between 1.75 and 1.9~keV dramatically improves $\chi ^2$ / d.o.f from 1.30 to 1.20.  

The best-fit model parameters for the six spectra are summarized in Table~\ref{table:fit}.
In light of the excessive local column density found in the 2007 \xmm\ observations,
the absorption in the model is allowed to exceed the nominal Galactic value of $N_H = 2 \times 10^{21}$~\cmii.
However, the slightly higher local column density detected in the 2007 \xmm\ spectra ($N_H = 2.7 \times 10^{21}$~\cmii , see Section~\ref{sec:xmm07}) is not required in other epochs, except perhaps in the lower-quality \asca\ spectrum where it is not as well constrained.
As discussed in Section~\ref{sec:xmm07}, this excess must be in our own galaxy (or at $z = 0$),
but why it varies remains unclear.
In the 2007 \xmm\ observations (top two panels in Fig.~\ref{abs_models}), the unabsorbed component dominates (i.e., low covering fraction) and the outflow, if at all present, covers not more than 20\% of the source; 
In the \suz, \asca, and \chandra\ spectra, it is the absorbed continuum that dominates (i.e., almost full coverage), while the model for the 2001 \xmm\ observation suggests an intermediate case in which about half of the source is covered by the absorber.
Thus, the variation of the partial covering fraction can explain much of the spectral variability observed in \pds. 
This would place the absorber at a distance from the source that is comparable to the source size.
Variable partial covering, however, can also be due to many small sources, e.g., an assortment of magnetic flares above an accretion disk, some of which are absorbed and some of which are not.
The size of the patchy absorber then would be of the accretion disk scale. 

Apparently, the column density in the absorption components also varies dramatically between epochs.
A column of a few times 10$^{22}$\,\cmii\ or higher can account for the spectral curvature between 2~-- 5~keV, which is observed in the 2001 \xmm, the \asca, and the \chandra\ spectra.
The \suz\ spectrum requires only the two high-ionization ($\log \xi > 3.5$) absorbers that produce the blueshifted Fe-K absorption lines \citep{reeves09} and that cover most of the source.
In the 2007 \xmm\ spectra, the alleged absorber does not significantly affect the spectrum up to 12~keV.
Indeed, the fits with absorption do not represent an improvement over those without absorption presented in Section~\ref{sec:xmm07}, despite the high column densities (but low covering fraction).
The absorption model in the 2001 \xmm\ observation requires (at least) two components,
and it needs to describe the absorption detected in the (simultaneous) \rgs\ spectrum of the same observation.
The most conspicuous absorption feature in that \rgs\ spectrum occurs at approximately 0.9~keV and can be discerned with the CCDs.
Such deep absorption lines require an absorber that essentially fully covers the source.
On the other hand, partial covering is required to produce the observed broad-band curvature of the spectrum.
This result was already reported by \citet{reeves03}.
Hence, the model proposed here for the 2001 \xmm\ observation, which is shown in Fig.~\ref{abs_models}, has a high-ionization absorber fully covering the source and a low ionization component only partially covering it ($\approx 50\%$), both with columns of a few times 10$^{22}$\,\cmii.
An additional component may be needed to fully account for the deep and blueshifted Fe-K edge observed at 7~-- 8~keV \citep[see Fig.~\ref{abs_models} and][]{reeves03}.

The reflection component is consistently present in all epochs.
Its relatively moderate flux indicates the reflector only partially intercepts the \x s from the primary power law source.
In fact, the models for all six epochs indicate it reflects about 20\% of the primary source (see also Section~\ref{sec:xmm07}).
This result is generally consistent with reflection off a moderately collimated outflow.
However, if the outflow is patchy as suggested by the varying partial covering,
the outflow segments can in fact be spread over a broader solid angle.
In the 2007 \xmm\ observations, the reflecting matter appears highly ionized ($\log \xi$ up to 4) and thus does not produce the soft \x\ emission that appears in all of the other epochs where it is approximately in the $2 \lsim  \log \xi \lsim 3$ range.
The reflection models tend to systematically prefer a super-solar Fe abundance, 
although the exact value is poorly constrained.
We deem such a high Fe abundance to be unphysical, and ascribe it mostly to the shortcomings of the model, rather than to an actual abundance effect.
Since in any case, without multiple identifiable narrow features it is impossible to expect a reliable assessment of abundances,
we fix the Fe abundance at three times solar (Section~\ref{sec:xmm07}).
The outflow velocities of the absorber/reflector are mostly constrained by the soft \x\ emission lines in the reflection component, which yields effective redshifts in most cases of $z = 0.10 - 0.14$.
If associated with an outflow in the reference frame of \pds\ ($z = 0.184$),
these redshifts translate into outflow velocities of $v_\mathrm{out}/c \approx$ --0.04 to --0.08.
Two exceptions to this velocity range are noted.
The first is the \suz\ spectrum in which the narrow absorption lines require a much faster outflow of $v_\mathrm{out}/c =$~--0.26 and --0.31 as reported by \citet{reeves09}.
The second exception is the 2001 \xmm\ observation that has a deep blueshifted edge-like feature at approximately 8~keV. 
When ascribed to the outflow, this edge implies $v_\mathrm{out}/c \approx$~--0.19 \citep{reeves03}.


Although the reflection component of the model does produce the observed flux below 1~keV, it does not always match all of the spectral details very well.
Consequently, at least formally, the fits are not always statistically acceptable. 
Indeed, invoking a partial covering absorber inserts additional uncertainty, as the fit is only partially sensitive to adding continuum and absorbing it.
Nonetheless, we suspect the deficient fits are likely due to shortcomings of the detailed models we use,
and do not necessarily invalidate the relevance of partial covering or reflection.
For example, even moderate broadening of discrete reflection features, due to velocity or to scattering, would make them less prominent.
Also, incomplete atomic data, the constant-density approximation, and the simplistic geometry used in the reflection model certainly affect the goodness of the fit.
Discrete features detected in the data, e.g. in the \xmm\ 2001 and \suz\ 2007 spectra, raise the confidence one has in the model, even if they do not result in superior statistical fits.
In fact, the 2001 \xmm\ spectrum is perhaps the hardest to fit and we find the partial covering model essential to even obtain an adequate description of the continuum.
In the following section we elaborate on the model for this observation that is derived from both the \pn\ and \rgs\ data.

\subsection{The 2001 RGS spectrum revisited}
\label{sec:rgs01}

\begin{figure}[pt!]
\begin{center}
\includegraphics[width=0.75\columnwidth, angle=-90]{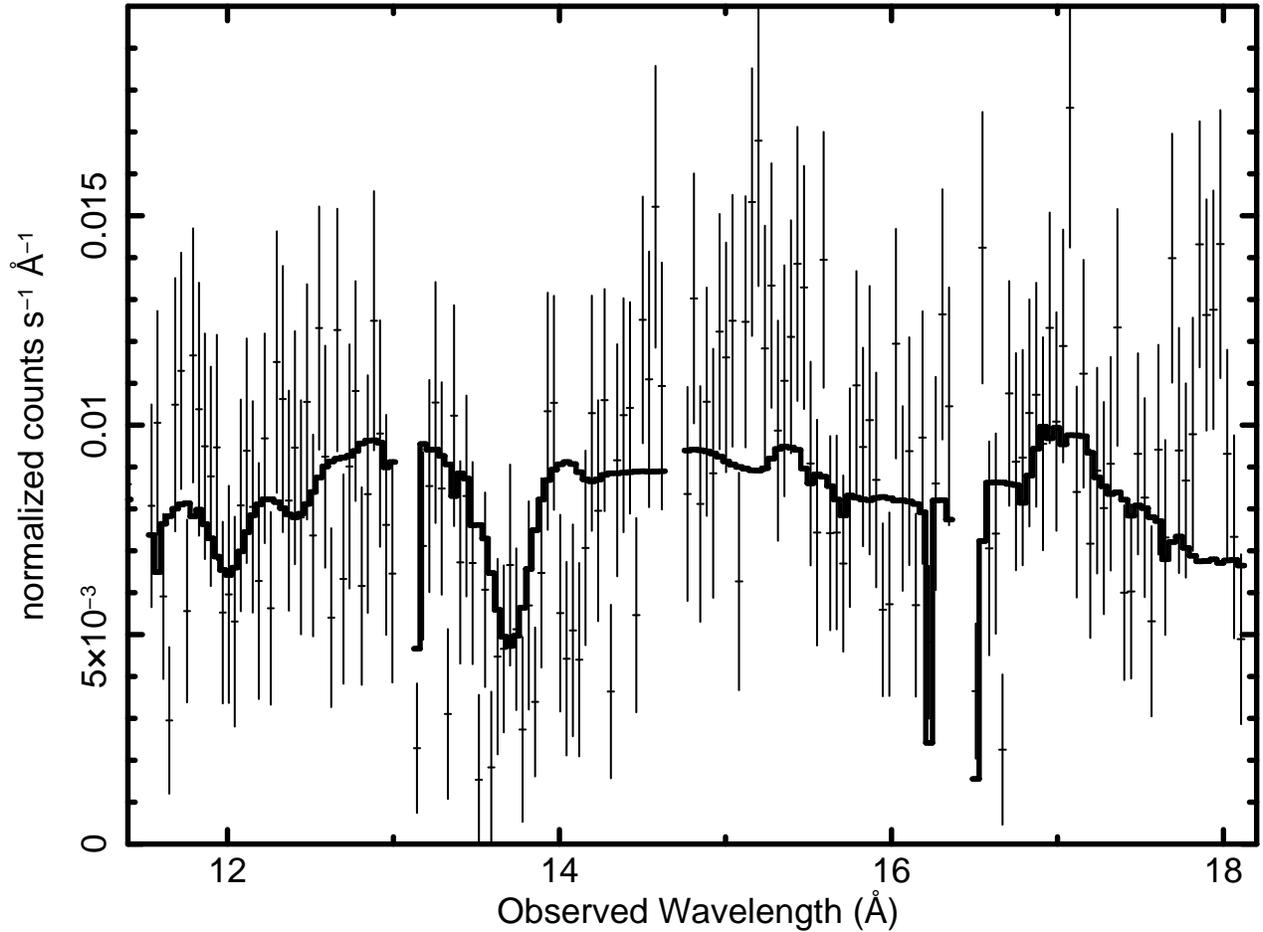} 
\caption{Extract form \rgs 2 spectrum of \pds\ obtained in 2001, and the proposed \xstar\ absorption model (solid line).
This model corresponds to $z = 0.12$ (an outflow velocity of 16,000 \kms\ in the quasar frame) that produces the most significant (redshifted) absorption feature at 13.7~\AA\ as a blend of Ne$^{+9}$ Ly\,$\alpha$ and Fe$^{+20}$ 2p -- 3d, the line at 12~\AA\ as the Fe$^{+23}$ 2s -- 3p doublet, and the  2p -- 3d emission line of Fe$^{+16}$ at 17~\AA\ from reflection.
}
\label{rgs01} 
\end{center}
\end{figure}

The 2001 \xmm\ observation is the only observation of \pds\ to provide identifiable absorption lines in the gratings. 
The least ambiguous absorption features in the 2001 \rgs\ spectrum are observed in the 11.5 -- 14.0~\AA\ range, and in particular there appears to be a deep and broad absorption trough between 13.5 -- 14.0~\AA.
We re-analyze the \rgs\ spectrum of \pds\ from the 2001 observation with the same absorption models as used above for the CCD data.
The method used here is not very different from that of \citet{reeves03}, who used \xstar\ tables.
We find that the ionization parameter of $\log \xi = 2.6$ (cgs units) given by \citet{reeves03} for the \rgs\ spectra
produces on top of the 14~\AA\ feature other strong lines that are not seen in the data.
In order to produce only a few features, we need to increase the ionization parameter to $\log \xi \approx 3.0$, to ensure that most species are fully ionized and only a trace of Fe-L remains.
Indeed, we are able to find a model with $\log \xi = 3.1$ that fits the \rgs\ absorption features,
and is marginally consistent with the fit to the \pn\ data obtained in Section~\ref{fit_results} (Table~\ref{table:fit}) that is not as sensitive to narrow line features.
The relevant part of the \rgs\ spectrum along with the model is presented in Fig.~\ref{rgs01}.
Unfortunately, \rgs 1 does not provide data between 11.5 -- 13.75~\AA, due to a failed chip.
In Fig.~\ref{rgs01}, thus, only \rgs 2 is shown, although both \rgs s are used in the fit and are consistent with each other.

According to the present model, several absorption features are shifted by $z = 0.13$ and can be tentatively identified in Fig.~\ref{abs_models}.
We identify the wide trough between 13.5 and 14.0~\AA\  predominantly with Ne$^{+9}$ Ly\,$\alpha$ ($\lambda _\mathrm{rest} = 12.134$~\AA) that is broad ($v_\mathrm{turb} \approx\ 2500~\kms$) and blended with the resonant ($f = 1.24$) 2p -- 3d line of  Fe$^{+20}$ ($\lambda _\mathrm{rest} = 12.284$~\AA).
The strength of the Ne$^{+9}$ Ly\,$\alpha$ line may indicate an overabundance of Ne, although no abundances could be reliably constrained.
The Ne$^{+9}$ Ly\,$\beta$ line in this model falls at 11.6~\AA\ ($\lambda _\mathrm{rest} = 10.239$~\AA), just next to a chip gap in \rgs 2 (and where \rgs 1 is still inactive), so it is not obvious in the spectrum.
Another relatively strong line is found at 12.01~\AA\ and identified as the Fe$^{+23}$ 2s -- 3p doublet ($\lambda _\mathrm{rest} = 10.621, 10.664$~\AA).
Finally, the model plotted in Fig.~\ref{rgs01} also predicts absorption by Fe$^{+22}$ at 12.4~\AA\  ($\lambda _\mathrm{rest} = 10.982, 11.021$~\AA), which is not clearly identified in the spectrum,
but consistent with the data uncertainty.
This blend could be overestimated in the model if the ionization parameter or the ionization balance are slightly off. 
These high charge states of Fe give an idea of the highly ionized state of the absorber.

The outflow velocity resulting from the aforementioned tentative line identifications as due to a shift of $z = 0.13$ is $v_\mathrm{out}/c = 0.13 - 0.184$, or $v_\mathrm{out} = -16,000~\pm\ 1,500~\kms$, slower than found by \citet{reeves03}, but consistent with the velocities seen for H Ly\,$\alpha$ absorption in the \uv.
The connection to the \uv\ absorber is puzzling since no high-ionization absorption lines were detected there \citep{obrien05}.
The determination of the outflow velocity obviously depends on the correct identification of the absorption lines.
Except for the 14~\AA\ feature, other identifications are marginally significant, and thus the outflow velocity derived mostly from one feature will always remain ambiguous.
The Fe-K edge observed with \pn\ at 8~keV still requires a high velocity of $\sim 50,000$~\kms.
This leaves the present velocity as well as that of \citet{reeves03} possible.
In any event, the CCD spectra that are of higher statistical quality and that were discussed in the previous section can not distinguish between the two interpretations of the \rgs\ data.
Fig.~\ref{rgs01} also highlights emission lines produced by the reflection model, which appear to be consistent with the \rgs\ data. The brightest line is at 17~\AA, which is the 2p -- 3d resonance line of Fe$^{+16}$ (redshifted with the outflow by $z$ = 0.13).  
Two additional narrow features that are not reproduced by the model appear in the \rgs 2 spectrum of Fig.~\ref{rgs01}, namely at 15.18~\AA\ and at 17.90~\AA.
The 15.18~\AA\ line appears in both \rgs s and corresponds to 13.43~\AA\ in the outflow frame, which could be the 1s -- 2p resonant line of Ne$^{+8}$.
The 17.9~\AA\ line is totally absent from the equally good spectrum of \rgs 1 (not shown) and can be discarded.

\section{Conclusions}
This paper presents the \xmm\ observation of \pds, which took place in September 2007,
 and attempts to generalize its results to the broader context of four other CCD-quality observations of the quasar acquired over the past decade, each one of them revealing a spectrum quite remarkably different from the others.
In a sense, the 2007 \xmm\ spectrum is the most straightforward to interpret, as no unambiguous absorption is present.
On the other hand, owing to the lack of ionized absorption, a significant reflection component can be discerned.
Global fits to the spectrum suggest it may be outflowing with respect to the quasar's rest frame, 
although this result is less robust that the mere existence of a reflecting medium.
Since an absorbing outflow has been identified in \pds,
a natural conjecture is to associate the reflector with outflow segments outside the line of sight.
The small solid angle coverage of the source of $\sim$~20\% found for the reflecting medium is consistent with this notion,
although the assumption of a single scatter, perfect reflector are clearly an over simplified picture of a more realistic reflecting outflow.
The velocities required to model the reflecting material are in the range of $v/c$ = --0.04 to --0.08, which is roughly comparable to the blueshifted absorption measured in the \uv\ and to the interpretation of the 2001 \rgs\ spectrum.  A separate component is needed to explain the much faster Fe-K absorption, observed e.g. with \suz.

The historical spectra of \pds\ point to a complex nature of the source and complex absorption.
We tested whether all spectra can be explained by a partial covering outflow model that both absorbs and reflects the central source and obtained mixed results.
This type of model is sufficiently flexible to produce a wide variety of spectra, but is only partially successful in reproducing those of \pds.
While the different over all spectral curvatures of \pds\ are produced by varying the fractional coverage of the source essentially from zero to unity, the spectral fits are not always statistically acceptable, and some narrow features predicted by the models are discrepant with the observed spectra.
These features vary in the data on time scales from days to years.  
The reflection component can explain some of the emission that is excessive of the power law continuum. In the 2007 epochs, it provides a good fit to the Fe-K emission, but in other epochs, its  details of the soft emission below 2~keV do not always match the data very well. 
Future \x\ instruments that would provide high S/N spectra on short time scales are imperative to further advance the study of such outflows.

\begin{acknowledgements} 
EB acknowledges funding from 
a NASA \xmm\ Guest Observer grant and from
NASA grant 08-ADP08-0076.
SK is supported at the Technion by the Kitzman Fellowship and by a
grant from the Israel-Niedersachsen collaboration program.
\end{acknowledgements}


\newpage

\newpage

%

\newpage

\begin{deluxetable}{lccccc}
\tabletypesize{ \footnotesize }
\tablecolumns{6} \tablewidth{0pt}
\tablecaption{
Observation Log (in arbitrary order)  
}
\tablehead{
   \colhead{Telescope} &
   \colhead{Instrument} &
   \colhead{Obs ID.} &
   \colhead{Start Date} &
   \colhead{Exposure (ks)}
}
\startdata
\xmm\ & \pn\ & 0501580101 (orbit 1421) & 2007 Sep. 12 & 91 \\
\xmm\ & \pn\ & 0501580201 (orbit 1422) & 2007 Sep. 14 & 88 \\
\xmm\ & \pn\ & 0041160101 (orbit 223) & 2001 Feb. 26 & 44 \\
\suz\ & XIS 0, 3 & 701056010  & 2007 Feb. 24 & 180 \\
\chandra\ & \hetg\ & 4063 & 2003 May 7 & 143 \\
\asca\ & SIS & 76078000& 1998 Mar. 7 & 43 \\
\enddata
\label{obs_log}
\end{deluxetable}

\newpage

\begin{deluxetable}{lcc}
\tabletypesize{ \footnotesize }
\tablecolumns{3} \tablewidth{0pt}
\tablecaption{
Model parameters obtained from fitting 2007 \pn\ spectra \tablenotemark{a}}
\tablehead{
   \colhead{Parameter} &
   \colhead{Orbit 1421} &
   \colhead{Orbit 1422}
}
\startdata
$N_H$ local (10$^{22}$\,\cmii)  &  \multicolumn{2}{c}{\multirow{1}{*}{0.267$^{+0.003}_{-0.002}$ }}  \\
POWER LAW \tablenotemark{b} & &  \\
\hspace{3pt} photon slope & 2.20 $\pm$ 0.01 & 2.25 $\pm$ 0.01 \\
\hspace{3pt} flux density at 1~keV (10$^{-3}\,$s$^{-1}$cm$^{-2}$keV$^{-1}$) & 4.0$^{+0.2}_{-0.1}$ & 2.4$^{+0.2}_{-0.1}$ \\
IONIZED REFLECTOR & & \\
\hspace{3pt} Fe abundance (solar units) & \multicolumn{2}{c}{\multirow{1}{*}{3.3 $^{+1.1}_{-0.6}$}} \\
\hspace{3pt} slope & \multicolumn{2}{c}{\multirow{1}{*}{tied to power-law slope}} \\
\hspace{3pt} redshift & \multicolumn{2}{c}{\multirow{1}{*}{0.12 $^{+0.03}_{-0.01}$}} \\
\hspace{3pt} ionization parameter $\xi$ (erg~s$^{-1}$cm) & \multicolumn{2}{c}{\multirow{1}{*}{10000$~^{+0}_{-4000}$} \tablenotemark{c} }\\
\hspace{3pt}  angular coverage $\Omega _\mathrm{refl}/ 2\pi$ \tablenotemark{d} & 0.19$ \pm$ 0.07 & 0.27$ \pm$ 0.10 \\
d.o.f. & \multicolumn{2}{c}{\multirow{1}{*}{2356}} \\
$\chi ^2$/d.o.f. & \multicolumn{2}{c}{\multirow{1}{*}{1.05}} \\
\enddata
\footnotesize 
\tablenotetext{a}{ Only one value is quoted for parameters that were tied for both data sets.
Errors are 90\% confidence limits.}
\tablenotetext{b}{ Redshifted uniformly to the \pds\ rest frame of $z$ = 0.184.}
\tablenotetext{c}{ Values greater than 10000 erg~s$^{-1}$cm are not available in the model.}
\tablenotetext{d}{Assuming no losses on the reflector.}
\label{tab:fit07}
\end{deluxetable}

\newpage

\begin{deluxetable}{lcccccc}
\tabletypesize{ \footnotesize }
\tablecolumns{7} \tablewidth{0pt}
\tablecaption{Best-fit parameters for partially covered absorption models with reflection}
\tablehead{
   \colhead{Parameter} &
   \colhead{\xmm} &
   \colhead{\xmm} &
   \colhead{\xmm} &
   \colhead{\suz} &
   \colhead{\asca}&
   \colhead{\chandra}  
\\  
   \colhead{ }&
   \colhead{\pn} &
   \colhead{\pn} &
   \colhead{\pn} &
   \colhead{XIS 0 \& 3} &
   \colhead{SIS} &
   \colhead{HETGS}  
\\ 
   \colhead{ }&
   \colhead{2007a} &
   \colhead{2007b} &
   \colhead{2001} &
   \colhead{2007} &
   \colhead{1998} &
   \colhead{2003} 
}

\startdata
$N_H$ local (10$^{22}\,$\cmii)  & 0.28$^{+0.02}_{-0.03}$ & 0.26$^{+0.02}_{-0.03}$ & 0.21$^{+0.02}_{-0.01}$ & 0.21$^{+0.02}_{-0.01}$  & 0.3$^{+0.1}_{-0.1}$ & 0.20$^{+0.1}_{-0.0}$  \\

POWER LAW \tablenotemark{a}& & & & & &\\
\hspace{3pt} photon slope &  \multicolumn{6}{c}{\multirow{1}{*}{fixed at 2.25}} \\
\hspace{3pt} 1~keV flux density & 
5.1 $\pm$ 0.3 & 2.5 $\pm$ 0.1 & 6.2  $\pm$ 0.2  & 3.1 $\pm$ 0.3 & 4.6 $\pm$ 0.9 & 3.2 $\pm$ 0.2  \\
\hspace{3pt} (10$^{-3}\,$s$^{-1}$cm$^{-2}$keV$^{-1}$) & & & & & & \\ 

IONIZED REFLECTOR & & & & & &\\
\hspace{3pt} photon slope & \multicolumn{6}{c}{\multirow{1}{*}{fixed to the primary power law}}  \\
\hspace{3pt} $\xi$ (erg~s$^{-1}$cm) & 10000 $^{+0}_{-2000}$ &  10000 $^{+0}_{-4000}$ &  1250 $^{+100}_{-100}$ &  1200 $^{+100}_{-150}$ &  850 $^{+300}_{-150}$ &  400 $^{+200}_{-100}$\\
\hspace{3pt} A$_{\mathrm Fe}$ (solar units) & \multicolumn{6}{c}{\multirow{1}{*}{fixed at 3.0}} \\
\hspace{2pt} ang. coverage $\Omega _\mathrm{refl}/ 2\pi$ \tablenotemark{b} & 0.12 $\pm$ 0.07 & 0.26 $\pm$ 0.10 & 0.14 $\pm$ 0.05 & 0.20 $\pm$ 0.07 & 0.08 $\pm$ 0.08 & 0.12 $\pm$ 0.12 \\


IONIZED ABSORBER \tablenotemark{c} & & & & & &\\
\hspace{3pt} $\log \xi ^\mathrm{high}$ &3.6 & \nodata & 3.0 $\pm$ 0.1 & $>$ 3.5 & 2.7 & 1.5\\
\hspace{3pt} $N_H^\mathrm{high}$ (10$^{22}\,$\cmii) & 
100 & \nodata &  2 $\pm$ 0.5 & 100 $\pm$ 30 & 100 $\pm$ 30 &  0.2 $\pm$ 0.2 \\
\hspace{3pt} $\log \xi ^\mathrm{low}$ & --2.7 & --1.7 & --1.5 & $>$ 3.5 & --3.2 & --1.8\\
\hspace{3pt} $N_H^\mathrm{low}$ (10$^{22}\,$\cmii) & 
100 & 100 & 6 $\pm$ 2 & 100 $\pm$ 30 & 3 $\pm$ 3& 8 $\pm$ 4  \\
\hspace{3pt} covering fraction & $< 0.2$ & $< 0.1$  & 0.5 $\pm$ 0.05 & 0.7 $\pm$ 0.3 & 0.9 $\pm$ 0.1 & 0.75 $\pm$ 0.05\\
$z$ (refl/abs) \tablenotemark{d} & 0.10~$\pm$~0.02 & 0.12~$\pm$~0.02 & 0.13 $\pm$ 0.02 \tablenotemark{e} & --0.08~$\pm$~0.02 \tablenotemark{f} & 0.11~$\pm$~0.02 & 0.14~$\pm$~0.02\\
$v_\mathrm{out}/c = z - 0.184$ \tablenotemark{d} & --0.08~$\pm$~0.02 & --0.06~$\pm$~0.02 & --0.05 $\pm$ 0.02 \tablenotemark{e} & --0.30~$\pm$~0.03 \tablenotemark{f} & --0.07~$\pm$~0.02 & --0.04~$\pm$~0.02\\
\cline {1-7}
d.o.f & 1281 & 1066 & 1053 & 214 & 105 & 275  \\
$\chi ^2$ / d.o.f & 1.10 & 0.99 & 1.18 & 1.30 & 1.00 & 1.29 \\
0.4--12 keV model flux &10.9 &  6.7 & 11.7 & 7.8 & 4.7 & 4.8 \\
(10$^{-12}$\,erg\,s$^{-1}$cm$^{-2}$) & &  & & & & \\
\enddata
\footnotesize 
\tablenotetext{a}{ Shifted to quasar rest frame at $z = 0.184$ } 
\tablenotetext{b}{Assuming no losses on the reflector.}
\tablenotetext{c}{ Fixed absorber properties: $v_\mathrm{turb} = 2,500~\kms$ and solar abundances.
Values without errors are not constrained.}
\tablenotetext{d}{ Common to reflector and absorber, except for \suz\ where absorber velocity is well constrained from lines. }
\tablenotetext{e} {For reflector and fully covering absorber}
\tablenotetext{f} {Independent redshift for the reflector of 0.06, or $v_\mathrm{out}/c = -0.12$.}
\label{table:fit}

\end{deluxetable}

\end{document}